\documentclass[10pt,aps,prd,fleqn,superscriptaddress,notitlepage,nofootinbib,preprintnumbers,nobalancelastpage]{revtex4-1}
\usepackage{amsmath,amssymb,graphicx,xspace,hyperref,todonotes,comment}
\usepackage[all]{hypcap}
\newcommand{\eps}{\varepsilon}

\hypersetup{hidelinks,
  pdfauthor={Stefan Hoeche,Mareen Hoppe,Daniel Reichelt},
  pdftitle={A simple algorithm for polarized parton evolution},
  pdfkeywords={Parton Shower,QCD Resummation}
}

\begin{document}
\preprint{FERMILAB-PUB-25-0595-T, CERN-TH-2025-165, MCNET-25-21}
\title{A simple algorithm for polarized parton evolution}
\author{Stefan~H{\"o}che}
\affiliation{Fermi National Accelerator Laboratory, Batavia, IL, 60510, USA}
\author{Mareen Hoppe}
\affiliation{Institute of Nuclear and Particle Physics, Technische Universit{\"a}t Dresden, 01062 Dresden, Germany}
\author{Daniel~Reichelt}
\affiliation{Theoretical Physics Department, CERN, CH-1211 Geneva, Switzerland}

\begin{abstract}
We present an algorithm to include the correlation between the production and decay 
planes of gluons in a parton-shower simulation. The technique is based on identifying the
charge currents responsible for the creation and annihilation of the vector field.
It is applicable in both the hard-collinear and the soft wide-angle region. As a function
of the number of particles, the algorithm scales linear in computing time and memory.
We demonstrate agreement with fixed-order perturbative calculations in the relevant kinematical
limits, and present a new observable that can be used to probe correlations beyond
current-current interactions.
\end{abstract}

\maketitle

\section{Introduction}
Parton showers have been an integral part of the collider physics program over the past four
decades~\cite{Webber:1986mc,Buckley:2011ms,Campbell:2022qmc}. They serve as computational tools
to implement the evolution of particles from high to low resolution scales, and are essential for
modeling the structure of the hadronic jets observed in experiments. They are needed to calibrate
the response of particle detectors and infer information about the hard interactions that are
used to probe the underlying theory of nature. In the context of previous measurements at the
Fermilab Tevatron and the CERN Large Hadron Collider (LHC), the precision of parton showers
has traditionally played an important, yet somewhat secondary role. While it was possible to
probe leading higher-order corrections, such as the functional form of the scale of running
coupling effects~\cite{Amati:1980ch}, and the two-loop effective coupling~\cite{Catani:1990rr},
it was difficult to resolve the precise radiation pattern at the fully differential level,
mainly because hadronization effects and detector capabilities were limiting the resolution.
This situation might change at the high-luminosity LHC, and is expected to change dramatically
at a potential Future Circular Collider (FCC)~\cite{FCC:2025lpp,FCC:2025uan,FCC:2025jtd}.
Despite a small dynamic range, the high statistics at the Tera-Z option of the FCC will
allow us to probe parton evolution in great detail. In addition, an improved understanding
of the perturbative QCD effects implemented by the parton shower will allow
for improved hadronization modeling, which is an important aspect in the determination
of the Higgs couplings at high precision. At the LHC, a large dynamic range will allow us
to further probe QCD evolution by measuring the detailed substructure of jets at highest energy
and highest experimental precision~\cite{Altheimer:2012mn,Larkoski:2017jix,Kogler:2018hem,Bonilla:2022wzp}.
The advances in experimental techniques will generally call for an increased precision in the
Monte-Carlo simulation of jet substructure through parton showers.

Among the many aspects of parton shower algorithms~\cite{%
  Webber:1983if,Bengtsson:1986gz,Bengtsson:1986et,Marchesini:1987cf,Andersson:1989ki,Bengtsson:1986hr,
  Webber:1987uy,Shatz:1983hv,Collins:1987cp,Knowles:1987cu,Knowles:1988vs,Knowles:1988hu,vanBeekveld:2022ukn,
  Nagy:2005aa,Nagy:2006kb,Schumann:2007mg,Giele:2007di,Platzer:2009jq,Hoche:2015sya,Fischer:2016vfv,
  Cabouat:2017rzi,Nagy:2012bt,Platzer:2012np,Nagy:2014mqa,Nagy:2015hwa,Platzer:2018pmd,Isaacson:2018zdi,
  Nagy:2019rwb,Nagy:2019pjp,Forshaw:2019ver,Hoche:2020pxj,DeAngelis:2020rvq,Holguin:2020joq,
  Gustafson:1992uh,Hamilton:2020rcu,Dasgupta:2018nvj,Dasgupta:2020fwr,
  Herren:2022jej,Assi:2023rbu,Hoche:2024dee,Preuss:2024vyu,Hoche:2017iem,Dulat:2018vuy,
  Gellersen:2021eci,FerrarioRavasio:2023kyg,vanBeekveld:2024qxs,
  Nagy:2008eq,Fischer:2017htu,Richardson:2018pvo,Karlberg:2021kwr,Hamilton:2021dyz},
the possibility to include polarization effects has been one of the earliest under
investigation~\cite{Webber:1987uy,Shatz:1983hv,Collins:1987cp,Knowles:1987cu,Knowles:1988vs,Knowles:1988hu}.
Its importance becomes clear when considering the analogue in classical electrodynamics:
To maximize the radio signal received on a dipole antenna, the receiver should be
co-polarized with the transmitter~\cite{Staelin:1994}. This is because
the polarization of the propagating electric field is linear in the case of a transmitting
dipole, and the direction is determined by the orientation of the emitter.
Such elementary correlations are essential to any gauge theory, hence they should be respected
by every simulation program. In recent years, the topic has received renewed interest~\cite{
  Nagy:2008eq,Fischer:2017htu,Richardson:2018pvo,Karlberg:2021kwr,Hamilton:2021dyz}.
The commonly employed algorithm to implement polarization effects is the Shatz-Collins-Knowles
method~\cite{Shatz:1983hv,Collins:1987cp,Knowles:1987cu,Knowles:1988vs,Knowles:1988hu}. It is based on
expressing the spin-dependent evolution of a particle in the hard collinear limit
through spin-density matrices that are determined from the spin-dependent decay matrix elements.
One of the algorithm's main drawbacks is that it requires modifications for soft wide-angle
gluon emission~\cite{Collins:1987cp,Hamilton:2021dyz}. Here, we propose a different, and much
simpler approach that does not have this undesirable feature. Making use of a recently developed
technique for the decomposition of the QCD splitting functions~\cite{Campbell:2025lrs}, we formulate
a correlation algorithm in terms of the most basic polarization information in the gluon field,
which stems from the orientation of the emitting QCD color dipole. We prove that, in the limit
of zero gluon virtuality, this is sufficient to account for all correlations in matrix elements with
up to three additional partons, and that corrections in matrix elements with four or more 
additional partons are strongly sub-leading. These subleading effects are true quantum interferences
at $\mathcal{O}(\alpha_s^3)$. Including them in the simulation only becomes necessary
once a similar accuracy is achieved in the evolution.
Based on this knowledge, we propose an observable that probes the co-polarization of the emitting
and absorbing QCD dipole antennae. This observable can be used to search for polarization effects
beyond those described by splitting functions up to $\mathcal{O}(\alpha_s^2)$.

The manuscript is organized as follows: Section~\ref{sec:polarization} reviews the computation
of the spin-dependent leading-order splitting functions, and derives the correlation algorithm.
Section~\ref{sec:results} presents a validation of the implementation in the Alaric
parton shower~\cite{Herren:2022jej} and discusses observable consequences.
In Sec.~\ref{sec:outlook} we discuss potential future developments.

\section{Polarized parton evolution}
\label{sec:polarization}
In this section, we derive a polarization correlation algorithm based on the
Lorentz structure of the splitting functions that define the branching probabilities
for a parton shower. It will be helpful to recompute those splitting functions
with a focus on their structure in terms of charge currents~\cite{Catani:1999ss,Campbell:2025lrs}.
This will be the topic of Sec.~\ref{sec:splitting_functions}. In Sec.~\ref{sec:algorithm}
the new expressions will be used to formulate the algorithm for the parton shower simulation.

\subsection{Splitting functions}
\label{sec:splitting_functions}
In order to obtain a compact form of the splitting functions, it is convenient
to use a Sudakov decomposition of the momenta~\cite{Sudakov:1954sw}. We would, however
like to obtain expressions that are sufficiently general to be used with various parton-shower
recoil schemes. This can be achieved in terms of a generic decomposition using forward
and generalized transverse momenta:
\begin{equation}\label{eq:sud_simple}
  \begin{split}
    p_1^\mu=&\;z_1\, p_{12}^\mu
    -\tilde{p}_{1,2}^\mu\;,
    \qquad\text{and}\qquad
    &p_2^\mu=&\;z_2\, p_{12}^\mu
    +\tilde{p}_{1,2}^\mu\;.
  \end{split}
\end{equation}
The individual components of this decomposition are given by
\begin{equation}\label{eq:def_z_ptilde}
    z_1=\frac{p_1\bar{n}}{p_{12}\bar{n}}\;,
    \qquad
    z_2=\frac{p_2\bar{n}}{p_{12}\bar{n}}\;,
    \qquad\text{and}\qquad
   \tilde{p}_{1,2}^\mu=z_1p_2^\mu-z_2p_1^\mu\;,
\end{equation}
where $p_{12}=p_1+p_2$, and where $\bar{n}$ is a light-like auxiliary vector,
which is linearly independent of the collinear direction. In this formulation
we can accommodate a broad class of kinematics mappings, two of which are
discussed in App.~\ref{sec:kinematics}.
In particular, we can show that certain components of the splitting functions
take particularly convenient forms in certain recoil schemes.

\subsubsection{Quark-to-quark splitting function}
\begin{figure}[t]
  \centerline{\includegraphics[width=.75\textwidth]{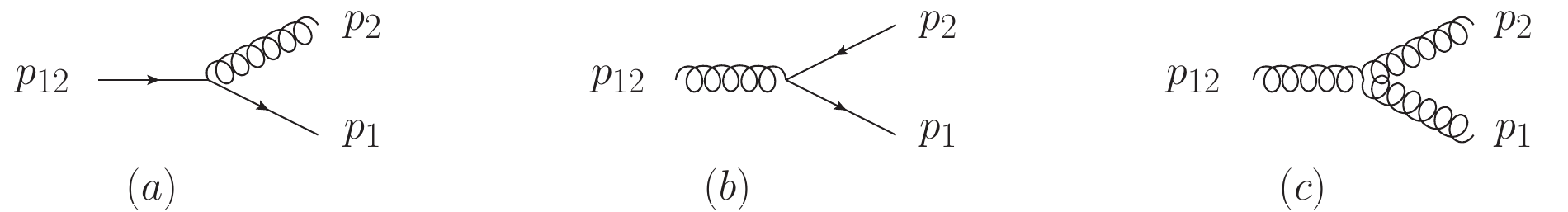}}
  \caption{Diagrams leading to the quark-to-quark-gluon splitting function~(a),
    the gluon-to-quark-antiquark splitting function~(b), and the
    gluon-to-gluon-gluon splitting function~(c) at tree level.
    The labels indicate the particle momenta referred to in the main text.
  \label{fig:splittings}}
\end{figure}
We begin the discussion with the quark splitting function. When working in
a physical gauge, it can be determined from the 2-particle fermion current,
which corresponds to the diagram in Fig.~\ref{fig:splittings}(a).
Here we consider massless partons only. The result with full spin dependence is
\begin{equation}\label{eq:coll_q_to_qg_1}
  \begin{split}
    P^{s'\bar{s}',\mu\nu}_{q\to q,\bar{s}s}(p_1,p_2)
    =&\;\frac{2C_F}{p_{12}^2}\,
    \frac{\bar{u}_{-\bar{s}}(\bar{n})\slash\!\!\!p_{12}\gamma^\mu u_{s'}(p_1)\,
    \bar{u}_{\bar{s}'}(p_1)\gamma^\nu\slash\!\!\!p_{12}u_{-s}(\bar{n})}{
    {\rm Tr}\big[\,\slash\!\!\!\bar{n}\slash\!\!\!p_{12}\,\big]}\;,
  \end{split}
\end{equation}
where the indices $s$ and $\bar{s}$ label the spin of the incoming fermion,
and $s'$ and $\bar{s}'$ label the spin of the outgoing fermion. We have used
Eq.~(3.4) of Ref.~\cite{Kleiss:1985yh} to define spinors for the fermion with
shifted momentum $\bar{p}_{12}^\mu=p_{12}^\mu-p_{12}^2/(2\bar{n}p_{12})\,\bar{n}^\mu$.
Due to the vector coupling of the gluon, Eq.~\eqref{eq:coll_q_to_qg_1}
is diagonal in the spin of the fermions, i.e.\ it is proportional
to $\delta_{\bar{s}}^{s'}\delta_{s}^{\bar{s}'}$. Fermion number conservation
implies that eventually the final-state fermion indices must be contracted
with $\delta_{s'\bar{s}'}$, thus reducing the fermion-spin dependence of
any product of $q\to qg$ splitting tensors to a simple identity matrix
in the spin indices of the fermion initiating the cascade~\cite{Somogyi:2005xz}.
Note that in the soft limit the spin of the radiating particle is generally
unaffected by the radiation. In order to implement polarization information
in $q\to qg$ splittings, we therefore only need to understand the gluon
polarization dependence of Eq.~\eqref{eq:coll_q_to_qg_1}.
Averaging over the redundant quark spins leads to the splitting tensor
\begin{equation}\label{eq:pqq_munu_sc}
  \begin{split}
    \langle P_{q\to q}^{\mu\nu}(p_1,p_2)\rangle
    =&\;\frac{C_F}{p_{12}^2}\,
    \frac{{\rm Tr}\big[\,\slash\!\!\!\bar{n}\slash\!\!\!p_{12}
    \gamma^\mu\slash\!\!\!p_1\gamma^\nu\slash\!\!\!p_{12}\,\big]}{
    {\rm Tr}\big[\,\slash\!\!\!\bar{n}\slash\!\!\!p_{12}\,\big]}
    =P_{\tilde{q}\to\tilde{q}}^{\mu\nu}(p_1,p_2)
    +\langle P^{{\rm(f)}\mu\nu}_{q\to q}(p_1,p_2)\rangle\;,
  \end{split}
\end{equation}
with its scalar and purely fermionic components
\begin{equation}\label{eq:pqq_munu_sc_components}
  \begin{split}
    P_{\tilde{q}\to\tilde{q}}^{\mu\nu}(p_1,p_2)
    =&\;\frac{C_F}{2}\,p_{12}^2\,S^\mu(p_1,p_2)S^\nu(p_1,p_2)\;,
    \qquad\text{and}\qquad
    &\langle P^{{\rm(f)}\mu\nu}_{q\to q}(p_1,p_2)\rangle
    =&\;\frac{C_F}{2}\bigg[\,-g^{\mu\nu}\,z_2
    -\frac{p_2^\mu p_2^\nu}{p_{12}^2}\,\bigg].
  \end{split}
\end{equation}
The scalar current, $S^\mu$, plays a central role in our algorithm. It is given by
\begin{equation}\label{eq:scalar_current}
  S^\mu(p_1,p_2)=\frac{(2p_1+p_2)^\mu}{p_{12}^2}\;.
\end{equation}
The explicit expressions for vertex factors resulting from this current
in specific kinematics are listed in App.~\ref{sec:scalar_emission}.

The conventional final-state summed splitting function is obtained by
contracting Eq.~\eqref{eq:pqq_munu_sc} with the polarization vectors of the
external gluon, $\varepsilon^\mu_{\pm}$, and summing over polarization states 
(see e.g.\ Ref.~\cite{Catani:1999ss}). The calculation is simplified
in a light-like axial gauge~\cite{
  Libby:1978bx,Ellis:1978sf,Ellis:1978ty,Bassetto:1982ma},
with the polarization vectors satisfying the relation:
\begin{equation}\label{eq:axial_gauge}
  \sum_{\lambda=\pm}\varepsilon^\mu_{\lambda}(p,\bar{n})
  \varepsilon^{\nu\,*}_{\lambda}(p,\bar{n})
  =d^{\mu\nu}(p,\bar{n})
  =-g^{\mu\nu}+\frac{p^\mu \bar{n}^\nu+p^\nu \bar{n}^\mu}{p\bar{n}}\;.
\end{equation}
Note that the gauge vector $\bar{n}^\mu$ does not have to be the same
as the auxiliary vector used to define the splitting kinematics in
Eq.~\eqref{eq:sud_simple}, but we will choose the two to be identical
for convenience. We obtain
\begin{equation}\label{eq:coll_q_to_qg_avg}
  \begin{split}
    \langle P_{q\to q}(p_1,p_2)\rangle
    =&\;\frac{C_F}{p_{12}^2}\,
    \frac{{\rm Tr}\big[\,\slash\!\!\!\bar{n}\slash\!\!\!p_{12}
    \gamma^\mu\slash\!\!\!p_1\gamma^\nu\slash\!\!\!p_{12}\,\big]}{
    {\rm Tr}\big[\,\slash\!\!\!\bar{n}\slash\!\!\!p_{12}\,\big]}\,
    d_{\mu\nu}(p_2,\bar{n})
    =P_{\tilde{q}\to\tilde{q}}(p_1,p_2)
    +\langle P_{q\to q}^{\rm(f)}(p_1,p_2)\rangle\;,
  \end{split}
\end{equation}
with the final-state summed scalar splitting function
and purely fermionic term~\cite{Campbell:2025lrs}
\begin{equation}\label{eq:coll_q_to_qg_avg_components}
  \begin{split}
    P_{\tilde{q}\to\tilde{q}}(p_1,p_2)=&\;C_F\frac{2z_1}{z_2}\;,
    \qquad\text{and}\qquad
    &\langle P_{q\to q}^{\rm(f)}(p_1,p_2)\rangle
    =&\;C_F\,(1-\eps)\,z_2\;.
  \end{split}
\end{equation}

\subsubsection{Gluon-to-quark splitting function}
The gluon-to-quark splitting function is determined by
the Feynman diagram in Fig.~\ref{fig:splittings}(b). The algebraic expression
is obtained using the methods of Ref.~\cite{Catani:1999ss}, and is given by
\begin{equation}\label{eq:coll_gqq_1}
  \begin{split}
    \,P^{s\bar{s},\bar{s}'s'}_{g\to q,\mu\nu}(p_1,p_2)
    =&\;\frac{T_R}{2p_{12}^2}\,d^\mu_{\;\rho}(p_{12},\bar{n})\,
    \bar{u}_{s}(p_1)\gamma^\rho u_{s'}(p_2)\,
    \bar{u}_{\bar{s}'}(p_2)\gamma^\sigma u_{\bar{s}}(p_1)\,
    d^\nu_{\;\sigma}(p_{12},\bar{n})\;.
  \end{split}
\end{equation}
As in the case of the $q\to qg$ splitting function, fermion number conservation
and the diagrammatic structure in the collinear limit imply that only the
fermion-spin diagonal terms contribute to the squared amplitude in the infrared
limits. We can therefore sum over the fermion spins and obtain the conventional
gluon-to-quark splitting function, which depends only on the gluon polarization:
\begin{equation}\label{eq:coll_gqq}
  \begin{split}
    \langle P^{\mu\nu}_{g\to q}(p_1,p_2)\rangle
    =&\;\frac{T_R}{2p_{12}^2}\,d^\mu_{\;\rho}(p_{12},\bar{n})
    {\rm Tr}[\,\slash\!\!\!p_1\gamma^\rho \slash\!\!\!p_2\gamma^\sigma\,]
    d^\nu_{\;\sigma}(p_{12},\bar{n})\\
    =&\;T_R\,\bigg[\,d^{\mu\nu}(p_{12},\bar{n})
    -p_{12}^2D^\mu(p_1,p_2)D^\nu(p_1,p_2)\,\bigg]\;.
  \end{split}
\end{equation}
The decay current, $D^\mu$, is defined as
\begin{equation}\label{eq:decay_current}
  \begin{split}
    D^\mu(p_1,p_2)=&\;d^{\mu\nu}(-p_{12},\bar{n})\,S_{\nu}(p_1,-p_{12})
    =\frac{d^\mu_{\;\nu}(p_{12},\bar{n})}{p_{12}^2}\,(p_1-p_2)^\nu\;.
  \end{split}
\end{equation}
We compute the explicit expressions for certain kinematics mappings
in App.~\ref{sec:scalar_decay}.
Using a standard Sudakov parametrization~\cite{Sudakov:1954sw},
taking the collinear limit, and summing over quark spins, we can write
Eq.~\eqref{eq:coll_gqq} in the familiar form of the spin-dependent
DGLAP splitting kernel~\cite{Catani:1999ss,Campbell:2025lrs}.
However, acknowledging the fact that Eq.~\eqref{eq:decay_current}
describes the absorption of the gluon on a scalar dipole radiator,
the structure of the splitting tensor is more obvious in the form
of Eq.~\eqref{eq:coll_gqq}, and we will therefore use this expression.

\subsubsection{Gluon-to-gluon splitting function}
The gluon-to-gluon splitting function is the most complicated of the
$1\to2$ splittings, due to the fact that there are three vector particles
that interact with each other. We find
\begin{equation}\label{eq:coll_ggg_step1}
  \begin{split}
    P^{\rho\sigma,\alpha\beta}_{g\to g,\mu\nu}(p_1,p_2)
    =&\;\frac{C_A}{2p_{12}^2}\,d_{\mu\lambda}(p_{12},\bar{n})\,
    \Gamma^{\rho\alpha\lambda}(p_1,p_2)\Gamma^{\sigma\beta\tau}(p_1,p_2)
      d_{\nu\tau}(p_{12},\bar{n})\;,
  \end{split}
\end{equation}
where $\Gamma^{\mu\nu\rho}(p,q)=g^{\mu\nu}(p-q)^\rho+g^{\nu\rho}(2q+p)^\mu-g^{\rho\mu}(2p+q)^\nu$
defines the Lorentz structure of the three-gluon vertex. The indices $\alpha$ and $\beta$
($\rho$ and $\sigma$) refer to the final-state gluon with momentum $p_2$ ($p_1$),
while the indices $\mu$ and $\nu$ refer to the initial-state gluon, which carries
momentum $p_{12}$.
We separate the resulting splitting tensor into a symmetric and an interference part
\begin{equation}\label{eq:coll_ggg}
  P_{g\to g,\mu\nu}^{\rho\sigma,\alpha\beta}(p_1,p_2)=
  P_{g\to g,\mu\nu,\rm(s)}^{\rho\sigma,\alpha\beta}(p_1,p_2)
  +P_{g\to g,\mu\nu,\rm(i)}^{\rho\sigma,\alpha\beta}(p_1,p_2)
  +P_{g\to g,\nu\mu,\rm(i)}^{\sigma\rho,\beta\alpha}(p_1,p_2)\;.
\end{equation}
The symmetric part reads
\begin{equation}\label{eq:coll_ggg_sym}
  \begin{split}
    P^{\rho\sigma,\alpha\beta}_{g\to g,\mu\nu,\rm(s)}(p_1,p_2)
    =\frac{C_A}{2}\,p_{12}^2\,\Big[\quad&
    S^\alpha(p_1,p_2)S^\beta(p_1,p_2)\,
    d_\mu^{\;\;\rho}(p_{12},\bar{n})d_\nu^{\;\;\sigma}(p_{12},\bar{n})\\
    +&S^\rho(p_2,p_1)S^\sigma(p_2,p_1)\,
    d_\mu^{\;\;\alpha}(p_{12},\bar{n})d_\nu^{\;\;\beta}(p_{12},\bar{n})\\
    +&D_\mu(p_1,p_2)D_\nu(p_1,p_2)\,g^{\rho\alpha}g^{\sigma\beta}\Big]\;.
  \end{split}
\end{equation}
while the asymmetric part is given by
\begin{equation}\label{eq:coll_ggg_asym}
  \begin{split}
    P^{\rho\sigma,\alpha\beta}_{g\to g,\mu\nu,\rm(i)}(p_1,p_2)
    =\frac{C_A}{2}\,p_{12}^2\,\Big[&
    -S^\rho(p_2,p_1)S^\beta(p_1,p_2)\,
    d_\mu^{\;\;\alpha}(p_{12},\bar{n})d_\nu^{\;\;\sigma}(p_{12},\bar{n})\\
    &+S^\rho(p_2,p_1)D_\nu(p_1,p_2)\,d_\mu^{\;\;\alpha}(p_{12},\bar{n})g^{\sigma\beta}
    -S^\beta(p_1,p_2)D_\mu(p_1,p_2)\,d_\nu^{\;\;\sigma}(p_{12},\bar{n})g^{\rho\alpha}\Big]\;.
  \end{split}
\end{equation}
We show in App.~\ref{sec:ggg_asym} that this interference contribution either vanishes
identically, or contributes at $\mathcal{O}(\alpha_s^3)$ and beyond.
This can be interpreted as an extension of the algorithm proposed in Ref.~\cite{Webber:1987uy}
to the case where $\langle\delta P\rangle$ is identically zero.
It is well known that four-gluon production also induces complicated sub-leading
color structures~\cite{Dokshitzer:1991wu,Gustafson:1992uh},
which have hitherto escaped implementation in standard parton showers.
We therefore find it justified to neglect the highly suppressed polarization effects
from Eq.~\eqref{eq:coll_ggg_asym} in our algorithm. If needed for practical applications
in (sub-)jet correlation measurements, one can make use of the fact that the jet multiplicity
is finite and typically rather small, which allows to implement the missing correlations through
existing matching and merging methods~\cite{Frixione:2002ik,Nason:2004rx,Frixione:2007vw,
  Alioli:2010xd,Hoche:2010pf,Hoeche:2011fd,Alwall:2014hca,Andre:1997vh,Catani:2001cc,Lonnblad:2001iq,
  Mangano:2001xp,Krauss:2002up,Lavesson:2007uu,Hoeche:2009rj,Hamilton:2009ne,Lonnblad:2011xx,
  Lonnblad:2012ng,Platzer:2012bs,Hoche:2019ncc,Lavesson:2008ah,Gehrmann:2012yg,Hoeche:2012yf,
  Frederix:2012ps,Lonnblad:2012ix,Hoche:2025gsb}.
Should the same terms have to be resummed without the remainder of the resummation reaching
$\mathcal{O}(\alpha_s^3)$ precision, one can still resort to the Shatz-Collins-Knowles
algorithm~\cite{Shatz:1983hv,Collins:1987cp,Knowles:1987cu,Knowles:1988vs,Knowles:1988hu}.

To conclude this section, we note that Eqs.~\eqref{eq:coll_gqq} and~\eqref{eq:coll_ggg}
are independent of the kinematics parametrization, as the invariants have not yet
been expressed in terms of transverse momenta, and the light-cone momentum fractions
are defined unambiguously by Eq.~\eqref{eq:def_z_ptilde}.

\subsubsection{Structure of the result}
\label{sec:structure}
Comparing Eqs.~\eqref{eq:pqq_munu_sc}, \eqref{eq:coll_gqq} and~\eqref{eq:coll_ggg_sym},
we find that most components of the splitting functions can be written as the scalar
radiation and decay vertices in Eqs.~\eqref{eq:scalar_current} and~\eqref{eq:decay_current}.
The only other components are the fermionic gluon production vertex,
$\langle P_{q\to q}^{{\rm(f)}\mu\nu}\rangle$, and the $d^{\mu\nu}$ component of the
gluon-to-quark splitting function. We therefore find precisely the pattern anticipated
in the introductory discussion: Correlations between gluon production and decay are
determined by simple dipole-dipole interactions that reflect the enhanced probability
to absorb energy from the vector field if the receiving antenna is co-polarized with
the emitter. Let us study this pattern in a bit more detail, using the most intricate
example of the gluon splitting function. Equation~\eqref{eq:coll_ggg_sym} consists of
three terms, each of which trivially factorizes into a component acting on the amplitude
and one acting on the complex conjugate amplitude:
\begin{equation}\label{eq:coll_ggg_sym_discussion}
  \begin{split}
    P^{\rho\sigma,\alpha\beta}_{g\to g,\mu\nu,\rm(s)}(p_1,p_2)
    =&\;\frac{C_A}{C_F}\Big[\,P_{\tilde{q}\to\tilde{q}}^{\alpha\beta}(p_1,p_2)\,
    d_\mu^{\;\;\rho}(p_{12},\bar{n})d_\nu^{\;\;\sigma}(p_{12},\bar{n})
    +P_{\tilde{q}\to\tilde{q}}^{\rho\sigma}(p_2,p_1)\,
    d_\mu^{\;\;\alpha}(p_{12},\bar{n})d_\nu^{\;\;\beta}(p_{12},\bar{n})\,\Big]\\
    &+\frac{C_A}{2}\,p_{12}^2\,D_\mu(p_1,p_2)D_\nu(p_1,p_2)\,g^{\rho\alpha}g^{\sigma\beta}\;.
  \end{split}
\end{equation}
After contraction with $g_{\alpha\beta}g_{\rho\sigma}$ we find that the
first (second) term describes the classical emission of gluon two (one) from gluon
one (two). The last term describes the absorption of the incoming gluon on the QCD
dipole formed by the two outgoing gluons. This indicates that the correct polarization
correlations will be implemented by an algorithm based on the following ideas:
\begin{itemize}
\item For radiation vertices, described by $P_{\tilde{q}\to\tilde{q}}^{\mu\nu}(p_i,p_j)$,
  store the polarization vector of the radiating QCD antenna. (We will discuss this vector
  in more detail in Sec.~\ref{sec:production_current}). Transfer the polarization information
  for the incoming parton with label $(ij)$ to the outgoing parton with label $i$.
\item For decay vertices, described by $D^\mu(p_i,p_j)$, form the polarization vector,
  $D^\mu$, and contract it with the polarization vector from the production vertex.
  Square the result, corresponding to the fact that one correlator exists for the
  amplitude, and one for the complex conjugate amplitude. Reweight the splitting
  probability accordingly.
\item For fermionic production vertices, do not store polarization information,
  corresponding to the fact that $\langle P_{q\to q}^{{\rm(f)}\mu\nu}(p_i,p_j)\rangle $ consists
  of a metric tensor, which will implement the trace of the decay tensor, and a term
  that is power suppressed in the strongly ordered limit.
\item For the term proportional to $d^{\mu\nu}$ in the fermionic decay of a gluon,
  implement the standard spin-averaged component of the splitting function.
\end{itemize}

\subsubsection{The gluon production current}
\label{sec:production_current}
In the previous section, we made use of the fact that, in the collinear limit,
the polarization vector for the production of a gluon is defined by a single
scalar radiation vertex of the form of Eq.~\eqref{eq:scalar_current}.
The extension beyond the limit is obtained with the help of the techniques in~\cite{Bassetto:1983mvz,Catani:1996vz}.
In this method, the one-gluon current for a QCD multipole reads
\begin{equation}\label{eq:scalar_current_sum}
  \begin{split}
    {\bf J}^\mu(\{p\};q)=&\;ig_s\sum_i\hat{\bf T}_i\,S^{\mu}(p_i,q)\;.
  \end{split}
\end{equation}
Here, $\hat{\bf T}$, are the charge operators, which are defined as
$(\hat{\bf T}^c_i)_{ab}=T^c_{ab}$ for quarks, $(\hat{\bf T}^c_i)_{ab}=-T^c_{ba}$
for anti-quarks, and $(\hat{\bf T}^c_i)_{ab}=if^{acb}$ for gluons.
Charge conservation implies that $\sum_i\hat{\bf T}_i=0$. If we work in the
color-flow representation~\cite{tHooft:1973alw,Maltoni:2002mq}, we can use the
relation $2\,if^{abc}\,T^a_{ij}T^b_{kl}=T^c_{kj}\delta_{il}-T^c_{il}\delta_{jk}$
to replace the charge operator of an adjoint with the two charge operators of
a bi-fundamental. This means simply that each gluon acts as both a QCD charge
and an anti-charge.
The scalar current can then be written as a double-sum over charge dipoles
\begin{equation}\label{eq:scalar_current_sum_dip}
  \begin{split}
    {\bf J}^\mu(\{p\};q)=&\;ig_s\,\frac{1}{n_q+n_g}
    \sum_{i\in{\rm q,g}}\sum_{k\in\bar{q},\bar{g}}\Big[\,
    \hat{\bf T}_i\,S^{\mu}(p_i,q)
    +\hat{\bf T}_k\,S^{\mu}(p_k,q)\,\Big]\;,
  \end{split}
\end{equation}
where $n_q$ and $n_g$ are the number of quarks and gluons in the multipole,
and where $g$ ($\bar{g}$) in the sums indicates that only the charge
(anti-charge) of the gluon is considered in the definition of $\hat{\bf T}_i$
($\hat{\bf T}_k$). In the leading-color approximation, the sums collapse to 
\begin{equation}\label{eq:scalar_current_sum_dip_lc}
  \begin{split}
    {\bf J}^\mu(\{p\};q)=&\;ig_s\,
    \sum_{i\in{\rm q,g}}\hat{\bf T}_i\,J_{i\bar{\imath}}^\mu(q)\,
    \qquad\text{where}\qquad
    J_{i\bar{\imath}}^\mu(q)=S^{\mu}(p_i,q)-S^{\mu}(p_{\bar{\imath}},q)\;.
  \end{split}
\end{equation}
Here the index $\bar{\imath}$ labels the parton which is color-connected to
parton $i$. At leading color, we can thus organize the calculation such that
there is always a relative minus sign between the scalar radiator for the charge
and the anti-charge. Dipole currents of this form obey the naive Ward identity,
and hence we find that the net effect of including the complete QCD multipole
in the splitting probability is that $S_\nu(p,q)d^{\mu\nu}(q,\bar{n})$ occurring
in the current correlators (see also App.~\ref{sec:scalar_emission}) is replaced
by $J_{i\bar{\imath}}^\mu(q)$ in Eq.~\eqref{eq:scalar_current_sum_dip_lc}.
This will form the basis for the assignment of polarization vectors in gluon
production in our algorithm.

\subsection{The correlation algorithm}
\label{sec:algorithm}
Using the ingredients above, we can formulate a Monte-Carlo algorithm that allows us
to easily implement polarization correlated evolution into any parton shower that
respects color coherence.
We assume that parton branching probabilities are separated into their scalar components
and collinear remainders. The details of this decomposition can be found, for example
in Refs.~\cite{Herren:2022jej,Assi:2023rbu,Hoche:2024dee}. We note that it can
straightforwardly be applied to any type of parton shower algorithm,
including dipole showers and antenna showers.

The correlation algorithm can be formulated as follows:
\begin{enumerate}
\item\label{algo:radiation}
  If a gluon is emitted using the scalar splitting function,
  store the polarization vector of the emitting color dipole.
  For a dipole that is formed by partons $i$ and $j$, the vector
  is given by the normalized scalar current\footnote{
    Note that we assume massless partons for simplicity.
    The extension to massive partons is straightforward
    and only involves replacing the scalar currents with
    their massive equivalent.}
  $j_p^\mu(p_i,p_j;q)=J_{ij}^\mu(q)/\sqrt{-J_{ij}^\nu(q)J_{ij,\nu}(q)}$.
  If gluon emission is described by the fermionic component of the
  $q\to qg$ splitting function, do not store a polarization vector.
\item\label{algo:decay}
  If a gluon undergoes a splitting according to the $\tilde{p}_{i,j}$-dependent
  part of the $g\to gg$ or $g\to q\bar{q}$ splitting function:
  \begin{itemize}
  \item If a polarization vector is defined for this gluon, contract it
  with the polarization vector for the decay, which is given by
  $j_d^\mu(p_i,p_j)=\tilde{p}_{i,j}^\mu/
        \sqrt{-\tilde{p}_{i,j}^\nu\tilde{p}_{i,j\,\nu}}$.
  Reweight the splitting probability accordingly. If the splitting is a
  $g\to gg$ transition, mark the two outgoing gluons as correlated,
  but do not define polarization vectors for their production.
  \item If no polarization vector is defined for this gluon,
  define the polarization vector for its correlated partner as the
  polarization vector of the decay, which is given by
  $j_d^\mu(p_i,p_j)=\tilde{p}_{i,j}^\mu/
        \sqrt{-\tilde{p}_{i,j}^\nu\tilde{p}_{i,j\,\nu}}$.
  \end{itemize}
\item\label{algo:pol_transfer}
  If a gluon emits another gluon, transfer the polarization vector and the
  information on the correlation partner from the incoming particle to the emitter.
\end{enumerate}
We note that this technique is much easier to implement than the Shatz-Collins-Knowles
algorithm, yet it provides most of the correlations that can occur in multi-parton
amplitudes, including the acausal quantum effects emulated by Step~\ref{algo:decay}.
It is straightforward to see that the computing time and memory associated with
the evaluation of the correlations is constant for each parton, therefore it scales
linearly with the number of particles in the final state. It is the most efficient
scaling possible for such an algorithm.

\section{Validation and Phenomenology}
\label{sec:results}
In this section, we will present the validation of our new algorithm, using 
some standard observables as well as a dedicated observable which is defined
with the help of the quantities in Sec.~\ref{sec:algorithm}. Let us first introduce
this observable. Based on the results in Sec.~\ref{sec:structure}, we find that
the correlations between production and decay of a gluon are described by the
current-current correlator $C_{i,j}^{1,2}=(j_p^\mu(p_i,p_j;p_{12})j_{d,\mu}(p_1,p_2))^2$,
which evaluates to
\begin{equation}\label{eq:current_correlator}
  \begin{split}
    C_{i,j}^{1,2}=&\;
    \frac{s_{i12}s_{j12}}{4s_{ij}s_{12}}\frac{(z_1+z_2)^2}{z_1z_2}
    \bigg[1+\frac{s_{12}}{s_{ij}}\frac{(s_{i1}-s_{j1}+s_{i2}-s_{j2})^2}{4s_{i12}s_{j12}}\bigg]^{-1}
    \bigg(\frac{s_{i1}-s_{i2}}{2s_{i12}}-\frac{s_{j1}-s_{j2}}{2s_{j12}}\bigg)^2\;.
  \end{split}
\end{equation}
In the double-soft limit, $p_1\to\lambda p_1$, $p_2\to\lambda p_2$, $\lambda\to0$,
this reduces to the azimuthal angle defined in Ref.~\cite{Dulat:2018vuy}, while in the
collinear limit, it reduces to the azimuthal angle of the squeezed correlator limit
defined in Ref.~\cite{Chen:2020adz}.\footnote{Ref.~\cite{Chen:2020adz} claims that the
  angular correlation pattern in the squeezed limit of the energy correlator
  provides a probe of the quantum structure of jet evolution. The discussion in
  Sec.~\ref{sec:polarization} and App.~\ref{sec:ggg_asym} shows that this claim only holds
  if one is able to resolve the impact of the finite remainder of higher-order virtual corrections,
  or the interference contributions of gluon splittings at $\mathcal{O}(\alpha_s^3)$
  and beyond. The remaining tree-level effects are of classical nature.}
We therefore propose $C_{i,j}^{1,2}$ as an observable to test polarization effects.
While Eq.~\eqref{eq:current_correlator} is difficult to implement experimentally,
it provides a theoretical tool to cleanly assess the various possible correlations
that occur in gluon production and subsequent decay processes.

Our tests will be performed with the Alaric final-state parton
shower~\cite{Herren:2022jej}, using the separation of kinematics mappings
detailed in Ref.~\cite{Assi:2023rbu}\footnote{The PyPy code for these tests
  can be found at~\url{https://gitlab.com/shoeche/pyalaric/-/tree/pol}.}.
As discussed in App.~\ref{sec:kinematics}, in particular Sec.~\ref{sec:scalar_decay},
the splitting kinematics used  for the collinear remainder functions is particularly
suited to describe the polarization effects in gluon decays.
We set $C_F=(N_c^2-1)/(2N_c)=4/3$ and $C_A=3$. All quarks are considered massless.
The running coupling is evaluated at two loop accuracy. We investigate $e^+e^-\to q\bar{q}$
at a center-of-mass energy of $\sqrt{s}=91.2$~GeV.

\begin{figure}[t]
  \centering
  \includegraphics[width=.475\textwidth]{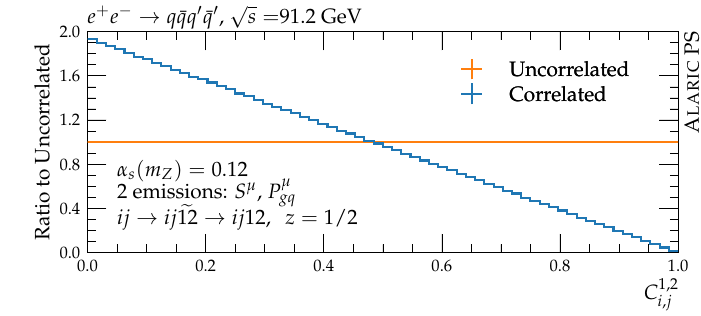}\hskip 2mm
  \includegraphics[width=.475\textwidth]{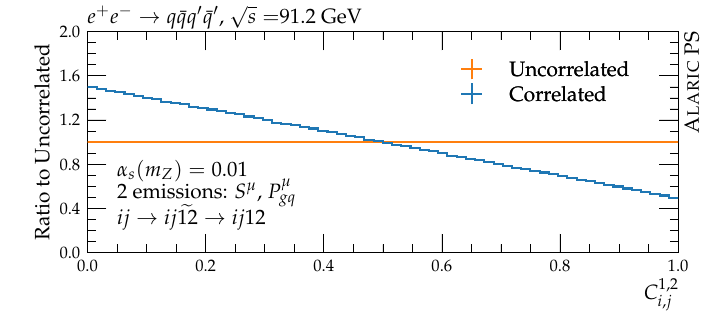}    
  \caption{The current-current correlator $C_{i,j}^{1,2}$ for the initial quark pair
    (labeled $i$ and $j$) and the emitted quark pair (labeled $1$ and $2$) in a 
    splitting sequence $ij\to ij\widetilde{12}\to ij12$ where the intermediate gluon
    decays to quarks. Polarization correlated evolution is compared to uncorrelated
    evolution. The left panel shows the result at $z=1/2$ in the second splitting.
    The right panel shows the complete result at $\alpha_s(m_Z)=0.01$.\label{fig:cij12_2em_q}}
\end{figure}
Figure~\ref{fig:cij12_2em_q} shows the current-current correlator $C_{i,j}^{1,2}$
for the initial quark pair (labeled $i$ and $j$) and an emitted quark pair
(labeled $1$ and $2$). The left panel shows the effect of polarized evolution
for $z=1/2$ in the second splitting. At this special kinematical point, polarization
effects on the gluon-to-quark splitting function, Eq.~\eqref{eq:coll_gqq}, are maximal.
The right panel shows the result for all values of $z$ at $\alpha_s=0.01$, where
kinematical edge effects are highly suppressed, which otherwise affect the admixture
of polarization-dependent and polarization-independent terms. In this case, the effect
of the polarized evolution is reduced to 50\%, in agreement with Eq.~\eqref{eq:coll_gqq}.

\begin{figure}[t]
  \centering
  \includegraphics[width=.475\textwidth]{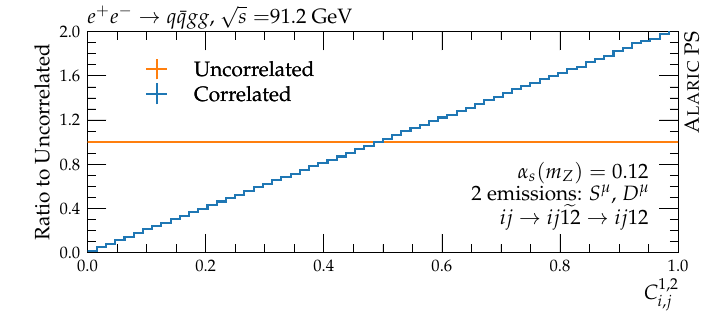}\hskip 2mm
  \includegraphics[width=.475\textwidth]{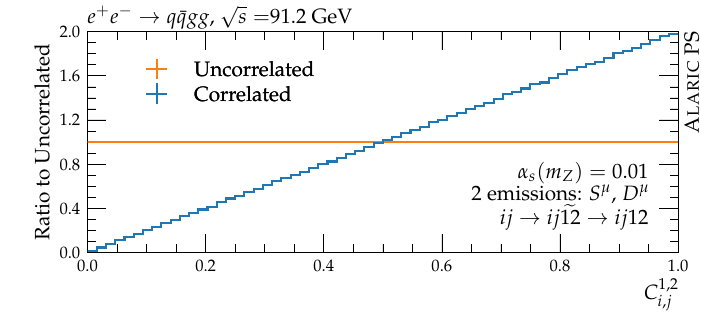}
  \caption{The current-current correlator $C_{i,j}^{1,2}$ for the initial quark pair
    (labeled $i$ and $j$) and the first two gluons (labeled $1$ and $2$) in a 
    splitting sequence $ij\to ij\widetilde{12}\to ij12$.
    Polarization correlated evolution is compared to uncorrelated evolution,
    and the second splitting is restricted to a decay, described by
    Eq.~\eqref{eq:decay_current}. We set $\alpha_s(m_Z)=0.12$ (left)
    and $\alpha_s(m_Z)=0.01$ (right).\label{fig:cij12_2em}}
\end{figure}
Figure~\ref{fig:cij12_2em} shows the current-current correlator $C_{i,j}^{1,2}$
for the initial quark pair (labeled $i$ and $j$) and the first two gluons
(labeled $1$ and $2$). We limit the evolution to only two emissions, in order
to obtain a clean identification of the individual partons. The splitting sequence
is $ij\to ij\widetilde{12}\to ij12$. We find the expected radiation pattern,
i.e. that the polarization correlated evolution generates a linear dependence
on $C_{i,j}^{1,2}$. The comparison between the left and the right figure shows
that using the current correlator as an observable, it is not necessary
to take the $\alpha_s\to0$ limit conventionally used to assess the correctness
of the parton shower in the region of vanishing recoil.

\begin{figure}[t]
  \centering
  \includegraphics[width=.475\textwidth]{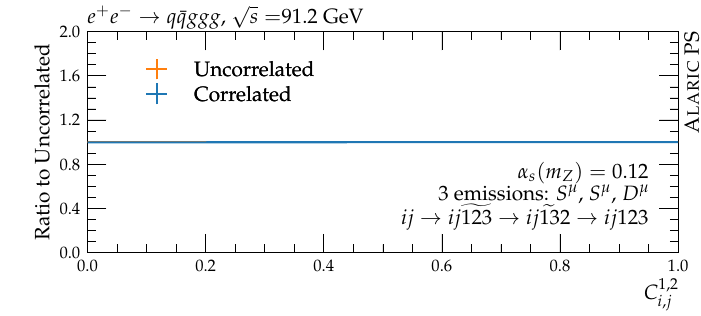}\hskip 2mm
  \includegraphics[width=.475\textwidth]{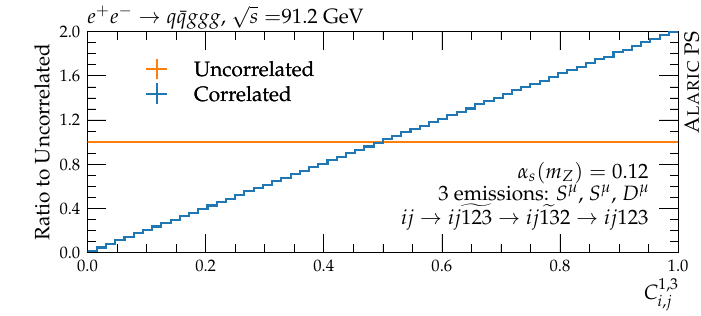}    
  \caption{The current-current correlators $C_{i,j}^{1,2}$ (left) and $C_{i,j}^{1,3}$ (right)
    in a splitting sequence $ij\to ij\widetilde{123}\to ij\widetilde{13}2\to ij123$.
    Polarization correlated evolution is compared to uncorrelated evolution,
    and the second splitting is restricted to scalar radiation, described by
    Eq.~\eqref{eq:scalar_current}, while the third is restricted to a decay,
    described by Eq.~\eqref{eq:decay_current}.\label{fig:cij12_cij13_3em}}
\end{figure}
Figure~\ref{fig:cij12_cij13_3em} shows the two current-current correlators
$C_{i,j}^{1,2}$ and $C_{i,j}^{1,3}$ in a configuration where we restrict
the parton shower to three emissions, and we constrain the second branching
to be a scalar emission, while the third is a decay. The splitting sequence
is $ij\to ij\widetilde{123}\to ij\widetilde{13}2\to ij123$. As expected,
the prediction of $C_{i,j}^{1,2}$ in this case is flat, even in the correlated
evolution algorithm, while $C_{i,j}^{1,3}$ has the behavior induced by
$(J_{ij}^\mu(p_{13})D_\mu(p_1,p_3))^2$. This validates in particular
the correct implementation of Step~\ref{algo:pol_transfer}
of the correlation algorithm in Sec.~\ref{sec:algorithm}. Note that small
deviations from the ideal pattern can occur due to kinematical effects
that stem from the radiation of the second gluon off the first. However,
these effects are practically invisible for only one intermediate emission,
even at $\alpha_s(m_Z)=0.12$.

\begin{figure}[t]
  \centering
  \includegraphics[width=.475\textwidth]{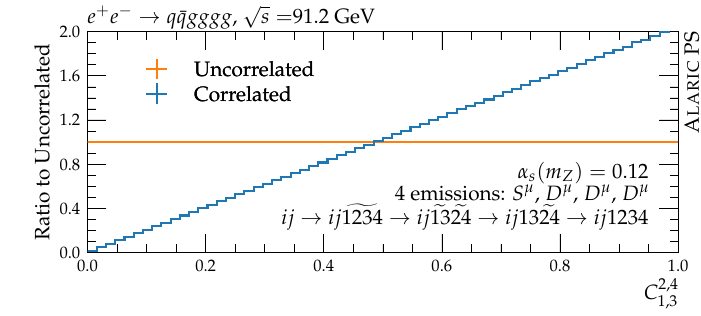}\hskip 2mm
  \includegraphics[width=.475\textwidth]{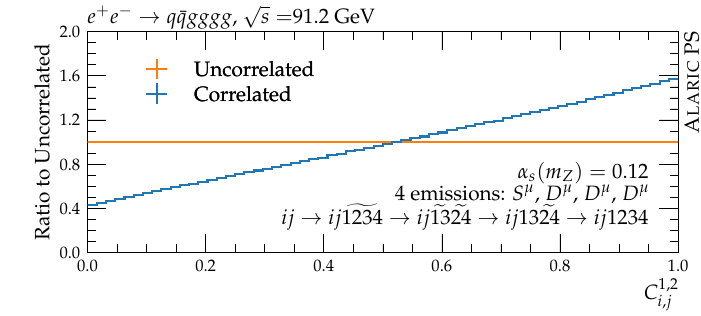}    
  \caption{The current-current correlators $C_{1,3}^{2,4}$ (left) and $C_{i,j}^{1,2}$ (right)
    in events with four branchings, where the second, third and fourth splitting are restricted 
    to a decay described by Eq.~\eqref{eq:decay_current}, and we enforce the splitting sequence
    $ij\to ij\widetilde{1234}\to ij\widetilde{13}\widetilde{24}\to ij12\widetilde{34}\to ij1234$.
    Decay probabilities are enhanced by a factor 100 to increase statistics.\label{fig:c1324_4em}}
\end{figure}
As a final test of the correct implementation of the correlation algorithm,
we investigate in Fig.~\ref{fig:c1324_4em} the current-current correlators
$C_{1,3}^{2,4}$ and $C_{i,j}^{1,2}$ in a configuration where we restrict
the number of splittings to four, and constrain the second, third and fourth
branching to the scalar decay in Eq.~\eqref{eq:scalar_current} in such a way
that the splitting sequence is given by $ij\to ij\widetilde{1234}\to 
  ij\widetilde{13}\widetilde{24}\to ij12\widetilde{34}\to ij1234$.
We find that the correlation between the decay planes of the two intermediate
gluons $\widetilde{13}$ and $\widetilde{24}$ is described nearly perfectly
by $(D_\mu(p_1,p_3)D^\mu(p_2,p_4))^2$, even for a standard value of $\alpha_s$.
This validates the correct implementation of both aspects of Step~\ref{algo:decay}
of the correlation algorithm in Sec.~\ref{sec:algorithm}. It is interesting
to note that $C_{i,j}^{1,2}$ still shows a strong correlation of
gluons $1$ and $2$, which stems from the initial gluon decay,
$ij\widetilde{1234}\to ij\widetilde{13}\widetilde{24}$.

\begin{figure}[t]
  \centering
  \includegraphics[width=.475\textwidth]{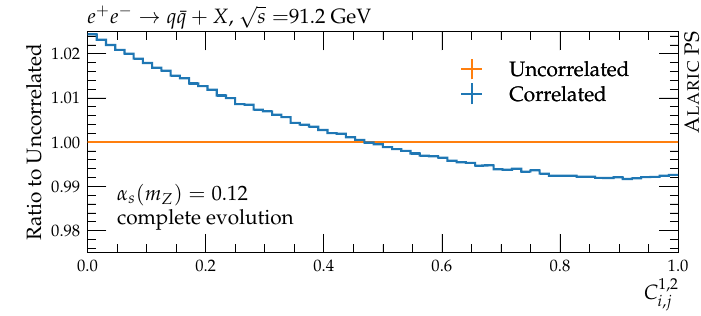}\hskip 2mm
  \includegraphics[width=.475\textwidth]{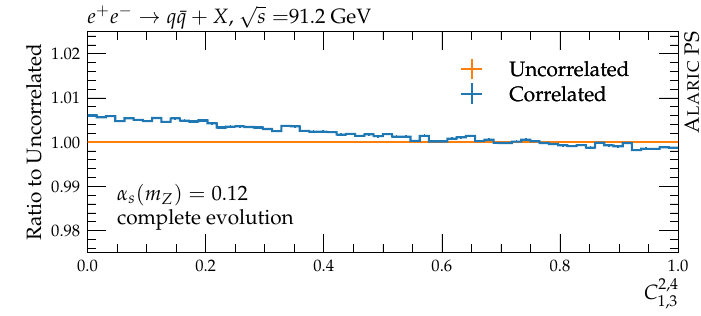}    
  \caption{The current-current correlators $C_{i,j}^{1,2}$ (left) and $C_{1,3}^{2,4}$ (right)
    in events generated with the full parton-shower evolution.\label{fig:correls_pheno}}
\end{figure}
Figure~\ref{fig:correls_pheno} shows the current-current correlators $C_{i,j}^{1,2}$ and
$C_{1,3}^{2,4}$ in events with the complete parton-shower evolution.
The large correlation effects observed in Figs.~\ref{fig:cij12_2em_q} and~\ref{fig:cij12_2em}
are strongly suppressed in this case, due to the much larger branching probability for scalar
emissions. This effect has previously been observed~\cite{Dulat:2018vuy}. It makes the
measurement of polarization effects in jet evolution very challenging. Note in particular
that the overall scale of correlation effects is at the same level as the expected perturbative
scale uncertainties, and the hadronization uncertainties in a typical simulation at LEP energies.
This was also pointed out in Ref.~\cite{Webber:1987uy}.

\begin{figure}[t]
  \centering
  \includegraphics[width=.495\textwidth]{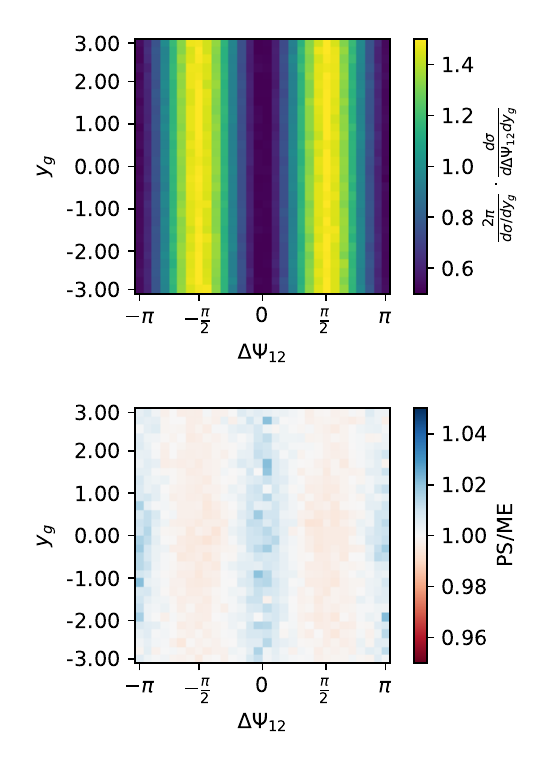}\hfill
  \includegraphics[width=.495\textwidth]{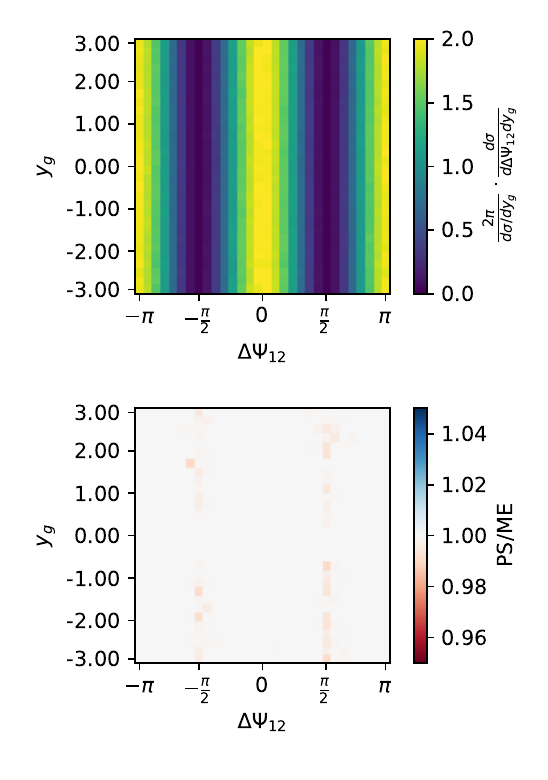}    
  \caption{Rapidity dependence of the azimuthal radiation pattern as defined in
    Ref.~\cite{Hamilton:2021dyz} at $\alpha_s=0.01$.
    Left: $e^+e^-\to q\bar{q}q'\bar{q}'$, compared to Eq.~(B.6) of
    Ref.~\cite{Ellis:1980wv}. Right: Decay component of $e^+e^-\to q\bar{q}gg$
    compared to the product of Eqs.~\eqref{eq:scalar_current_sum_dip_lc}
    and~\eqref{eq:decay_current}.
    See the main text for details.\label{fig:y_dependence}}
\end{figure}
In Fig.~\ref{fig:y_dependence} we show the rapidity dependence of the azimuthal radiation
pattern that was investigated in Ref.~\cite{Hamilton:2021dyz}. We limit the evolution to
two emissions. The azimuthal angle $\psi_{12}$ is defined as the angle between the production
and decay planes of the intermediate gluon generated in the first emission, using the
emitting quark and the more energetic of the gluon decay products to define the leading
directions. In the left panels, we show the radiation pattern generated by the parton
shower in $e^+e^-\to q\bar{q}q'\bar{q}'$, on the right we show the radiation pattern in
$e^+e^-\to q\bar{q}gg$, but only including the scalar gluon production current, and
only the decay component on the third line of Eq.~\eqref{eq:coll_ggg_sym}.
In the lower left panel, we compare the parton-shower result to Eq.~(B.6)
of Ref.~\cite{Ellis:1980wv}. In the lower right panel, we compare it to
$(J_{ij}^\mu(p_{12})D_\mu(p_1,p_2))^2$, where $p_1$ and $p_2$ are the momenta
of the two gluons, and $p_i$ and $p_j$ are the momenta of the quarks.
In both cases we find very good agreement.

In Ref.~\cite{Hamilton:2021dyz} it was observed that, depending on the precise
implementation of the parton shower, the Shatz-Collins-Knowles algorithm for spin correlations
may not capture the required azimuthal modulation at small rapidity. We observe no such
effects in Fig.~\ref{fig:y_dependence}, confirming that our new algorithm respects the
polarization information, independent of other phase-space variables. For this, it is
important that the information on the production current is appropriately Lorentz
transformed during any subsequent splitting.

\section{Outlook}
\label{sec:outlook}
For dipole antennae, the angular correlation between emitter and receiver determines
the power transfer through electromagnetic waves. Being a consequence of Maxwell's
equations~\cite{Maxwell:1865zz}, the effect is omnipresent in gauge theories
and also determines certain angular correlations that occur in QED and QCD
particle evolution during highly energetic reactions at collider experiments.
These polarization effects should be respected in any simulation of high-energy
collisions by Monte-Carlo event generators. In this manuscript, we have proposed a novel
algorithm to include polarization information in parton-showers. The method
scales linearly in computing time and memory, and is thus one of the most efficient techniques
available. Its true advantage does, however lie in its straightforward implementation and 
interpretation, and in the fact that the results provide a new observable to probe correlations
beyond the angular radiation patterns typically tested in the context of polarized
parton evolution. In the future, we will investigate the connection to the helicity formalism,
and develop the matching to existing perturbative calculations for LHC and FCC physics.

\section*{Acknowledgments}
We would like to thank Frank Siegert for many inspiring discussions on
spin correlations and their simulation in Monte-Carlo event generators.
This manuscript has been authored by Fermi Forward Discovery Group, LLC
under Contract No. 89243024CSC000002 with the U.S.\ Department of Energy,
Office of Science, Office of High Energy Physics. This research used resources
of the National Energy Research Scientific Computing Center (NERSC),
a Department of Energy Office of Science User Facility using NERSC award ERCAP0028985.
The work of S.H. was supported by the U.S. Department of Energy,
Office of Science, Office of Advanced Scientific Computing Research,
Scientific Discovery through Advanced Computing (SciDAC-5) program,
grant ``NeuCol''. D.R.\ is supported by the European Union under the HORIZON
program in Marie Sk{\l}odowska-Curie project No. 101153541. 
The work of M.H. was partly supported by a doctoral scholarship from the
Friedrich-Ebert-Stiftung. M.H. thanks the Fermilab Particle Theory Department
for hospitality during part of the work.
We thank the CERN Theoretical Physics Department for hospitality
during the 2025 conference on Parton Showers and Resummation,
where we first presented this algorithm.

\appendix
\section{Kinematics}
\label{sec:kinematics}
\begin{figure}[t]
  \centerline{\includegraphics[width=0.9\textwidth]{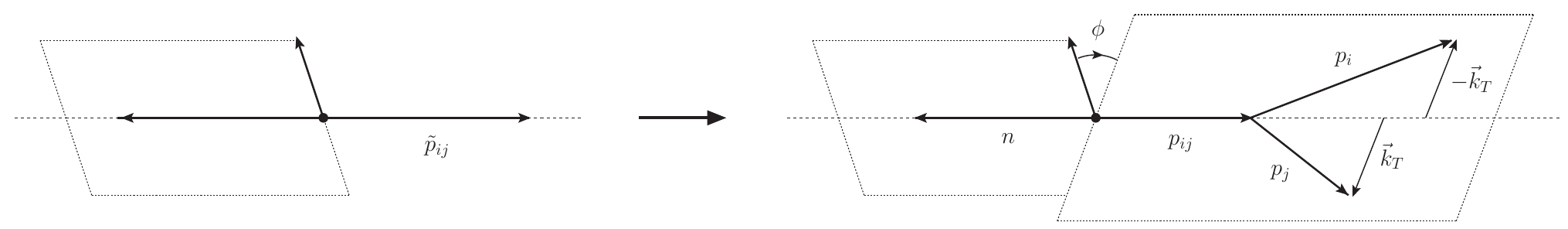}}
  \centerline{\includegraphics[width=0.9\textwidth]{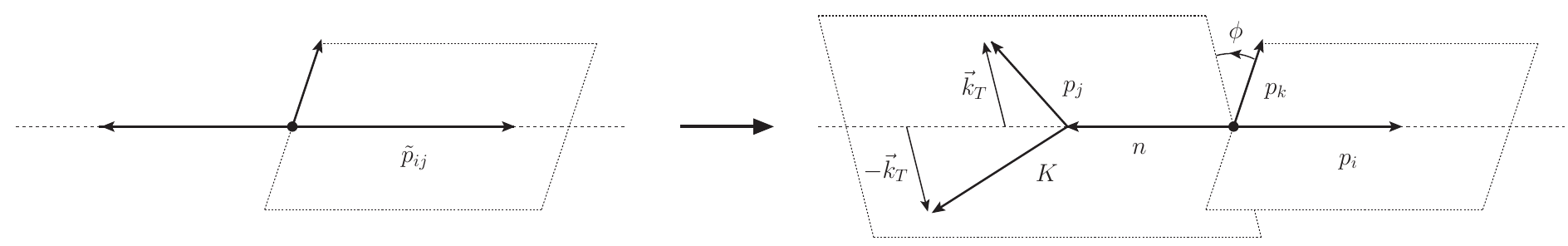}}
  \caption{Top: The splitting kinematics, described by Eq.~\eqref{eq:sud_general} and the choice $\alpha=1$.
    The momentum $n$ takes the recoil in the splitting, and the momentum $p_k$ defines
    a reference direction for the azimuthal angle.
  Bottom: The radiation kinematics, described by Eq.~\eqref{eq:sud_general} and the choice $\alpha=0$.
    The momentum $K$ takes the recoil in the splitting, and the momentum $p_k$ defines
    a reference direction for the azimuthal angle.
    \label{fig:kinematics}}
\end{figure}
In order to obtain a concrete form of the splitting functions, one must define a Sudakov
decomposition of the final-state momenta. Ideally, this parametrization will be designed
to reflect the dynamics of the radiation and decay vertices.
A generic decomposition in terms of forward and backward light-cone momenta is achieved as follows:
\begin{equation}\label{eq:sud_general}
  \begin{split}
    p_i^\mu=&\;z_i\tilde{p}_{ij}^\mu
    +\frac{\alpha^2 {\rm k}_\perp^2+p_i^2}{z_i\,2\tilde{p}_{ij}\bar{n}}\,
    \bar{n}^\mu-\alpha k_\perp^\mu\;,
    \qquad\text{and}\qquad
    &p_j^\mu=&\;z_j\tilde{p}_{ij}^\mu
    +\frac{{\rm k}_\perp^2+p_j^2}{z_j\,2\tilde{p}_{ij}\bar{n}}\,
    \bar{n}^\mu+k_\perp^\mu\;,
  \end{split}
\end{equation}
where $k_\perp \tilde{p}_{ij}=k_\perp\bar{n}=0$ and ${\rm k}_\perp^2=-k_\perp^2$.
The light-cone momentum fractions satisfy $z_i+z_j=1$ and are given by Eq.~\eqref{eq:def_z_ptilde}.
The parameter $\alpha$ depends on the type of splitting: $\alpha=1$ corresponds to
$k_\perp$--symmetric splitting kinematics where final states are treated on equal footing,
while $\alpha=0$ corresponds to radiation kinematics, where particle $i$ is identified.
The two cases are sketched in Fig.~\ref{fig:kinematics}.
The vector $n$ shown in the figure is related to $\bar{n}$
through an on-shell shift. We define the forward light-cone direction as
\begin{equation}\label{eq:sud_pijt}
    \tilde{p}_{ij}^\mu=\frac{p_i^\mu+\alpha p_j^\mu}{z_i+\alpha z_j}
    -\frac{(p_i+\alpha p_j)^2}{z_i+\alpha z_j}
    \frac{\bar{n}^\mu}{2(p_i+\alpha p_j)\bar{n}}
    =p_{ij}^\mu-\frac{1-\alpha}{z_i+\alpha z_j}\,\tilde{p}_{i,j}^\mu
    -\frac{(p_i+\alpha p_j)^2}{z_i+\alpha z_j}
    \frac{\bar{n}^\mu}{2(p_i+\alpha p_j)\bar{n}}\;,
\end{equation}
The transverse three-momentum squared in a frame where $\tilde{p}_{ij}$
and $\bar{n}$ are back-to-back is related to the invariants by
\begin{equation}\label{eq:kt_as_func_of_pij}
    {\rm k}_\perp^2=\frac{z_i z_j p_{ij}^2\,\kappa_{i,j}^2}{(z_i+\alpha z_j)^2}\;,
    \qquad\text{where}\qquad
    \kappa_{i,j}^2=1-\frac{z_j p_i^2+z_i p_j^2}{z_i z_j p_{ij}^2}\;.
\end{equation}

\subsection{Radiation off a scalar}
\label{sec:scalar_emission}
Here we compute the vertex factor associated with gluon radiation off the scalar vertex
defined in Eq.~\eqref{eq:scalar_current}. In the collinear limit, the produced gluon
is associated with a polarization tensor that reflects either its propagation, or the
sum over final-state helicities. The general expression for the product of a scalar
emission vertex and this polarization sum is given by
\begin{equation}\label{eq:sq_to_sqg}
    S^\rho(p_i,p_j)\,d^\mu_{\;\rho}(p_j,\bar{n})=\frac{1}{p_{ij}^2}
    \frac{2}{z_j}\,\bigg[\,\tilde{p}_{i,j}^\mu
    +(p_{ij}^2-p_i^2)\frac{\bar{n}^\mu}{2p_{ij}\bar{n}}\bigg]\;.
\end{equation}
It is interesting to note that the soft enhancement $1/z_j$ already appears at vertex level,
which is a consequence of the gluon propagator in axial gauge, Eq.~\eqref{eq:axial_gauge}.
Using the Sudakov decomposition, Eq.~\eqref{eq:sud_general} we find
\begin{equation}\label{eq:sq_to_sqg_gk}
  \begin{split}
    S^\rho(p_i,p_j)\,d^\mu_{\;\rho}(p_j,\bar{n})=&\;
    \frac{1}{p_{ij}^2}\frac{2}{z_j}\,\bigg[\,(z_i+\alpha z_j)k_\perp^\mu+
    \bigg(\frac{z_i+\alpha z_j}{z_j}\,{\rm k}_\perp^2
    +\frac{1+z_i}{2z_j}\,p_j^2\bigg)\,
    \frac{\bar{n}^\mu}{p_{ij}\bar{n}}\bigg]\;.
  \end{split}
\end{equation}
In radiation kinematics, $\alpha=0$, we obtain
\begin{equation}\label{eq:zipjmzjpi_rad}
  \begin{split}
    S^\rho(p_i,p_j)\,d^\mu_{\;\rho}(p_j,\bar{n})=&\;\frac{1}{p_{ij}^2}
    \frac{2z_i}{z_j}\bigg[\,k_\perp^\mu+\bigg({\rm k}_\perp^2
    +\frac{1+z_i}{2z_j}\,p_j^2\bigg)\frac{\bar{n}^\mu}{p_j\bar{n}}\bigg]\;.
  \end{split}
\end{equation}
In the collinear limit, $k_\perp\to\lambda k_\perp$, and for massless final
states, this is the expected polarized splitting amplitude
\begin{equation}\label{eq:zipjmzjpi_rad_coll}
  \begin{split}
    \lambda\,S^\rho(p_i,p_j)\,d^\mu_{\;\rho}(p_j,\bar{n})
    \to&\;2\sqrt{\frac{1}{p_{ij}^2}\frac{z_i}{z_j}}\,
    \frac{k_\perp^\mu}{{\rm k}_\perp}
    +\mathcal{O}(\lambda)\;.
  \end{split}
\end{equation}

\subsection{Vector decay to scalars}
\label{sec:scalar_decay}
Here we compute the vertex factor associated with the gluon decay in Eq.~\eqref{eq:decay_current}.
We find the general expression
\begin{equation}\label{eq:g_to_sqsq}
    D^\mu(p_i,p_j)=\frac{2}{p_{ij}^2}\,
    \bigg[\,\tilde{p}_{i,j}^\mu+(p_i^2-p_j^2)\frac{\bar{n}^\mu}{2p_{ij}\bar{n}}\bigg]\;.
\end{equation}
Using the Sudakov decomposition in Eq.~\eqref{eq:sud_general}, we obtain
\begin{equation}\label{eq:g_to_sqsq_sd_rad}
    D^\mu(p_i,p_j)=\frac{2}{p_{ij}^2}\,
    \bigg[\,(z_i+\alpha z_j)k_\perp^\mu+
    \bigg(\frac{z_i-\alpha z_j}{z_i+\alpha z_j}\,p_{ij}^2
    -\frac{2(1-\alpha)}{z_i+\alpha z_j}\,(z_j p_i^2+z_i p_j^2)\bigg)\,
    \frac{\bar{n}^\mu}{2p_{ij}\bar{n}}\bigg]
\end{equation}
In splitting kinematics, $\alpha=1$, Eq.~\eqref{eq:g_to_sqsq_sd_rad} reduces
to an expression, which reflects the symmetry of the process
\begin{equation}\label{eq:g_to_sqsq_sd_split}
  \begin{split}
    D^\mu(p_i,p_j)=&\;\frac{2}{p_{ij}^2}\,
    \bigg[\,k_\perp^\mu+(z_i-z_j)\,\frac{p_{ij}^2}{2p_{ij}\bar{n}}\,\bar{n}^\mu\bigg]\;.
  \end{split}
\end{equation}
In the collinear limit, $k_\perp\to\lambda k_\perp$, and for massless final
states, this is the expected polarized splitting amplitude
\begin{equation}\label{eq:g_to_sqsq_sd_split_coll}
  \begin{split}
    \lambda\,D^\mu(p_i,p_j)\to
    2\sqrt{\frac{\vphantom{p_{ij}^2}z_iz_j}{p_{ij}^2}}\,
    \frac{k_\perp^\mu}{{\rm k}_\perp}+\mathcal{O}(\lambda)\;.
  \end{split}
\end{equation}

\subsection{Polarization transfer}
\label{sec:pol_transfer}
For gluon splitting functions we also need to understand how the polarization of
the incoming gluon is affected when radiating a new gluon. The relevant expressions
are given by the product of two polarization tensors with different momenta:
\begin{equation}\label{eq:pol_transfer}
  \begin{split}
  d^{\mu\rho}(p_{ij},\bar{n})d_\rho^{\;\;\nu}(p_i,\bar{n})=&\;
  -d^{\mu\nu}(p_{ij},\bar{n})+\bigg[\,\tilde{p}_{i,j}^\mu
  +(p_{ij}^2+p_i^2-p_j^2)\frac{\bar{n}^\mu}{2p_{ij}\bar{n}}\bigg]\frac{\bar{n}^\nu}{p_i\bar{n}}\;,
  \end{split}
\end{equation}
In the collinear limit, $k_\perp\to\lambda k_\perp$, and for massless final
states, this simply gives the original polarization tensor
\begin{equation}\label{eq:pol_transfer_coll}
  d^{\mu\rho}(p_{ij},\bar{n})d_{\rho\nu}(p_i,\bar{n})
  \to -\,d^{\mu\nu}(p_{ij},\bar{n})+\mathcal{O}(\lambda)\;.
\end{equation}

\section{Interference contribution to the gluon splitting tensor}
\label{sec:ggg_asym}
In order to form a complete amplitude for the production and evolution of a gluon,
the splitting tensor in Eq.~\eqref{eq:coll_ggg_step1} must eventually be contracted with
other Lorentz structures involving the polarization tensors of the gluons with momentum
$p_1$ and $p_2$. In the strongly ordered soft and collinear limit, and at leading power,
these tensors become projectors, cf.\ App.~\ref{sec:pol_transfer}.
Replacing $d^{\mu\nu}(p,\bar{n})\to d^{\mu\rho}(p,\bar{n})d_\rho^{\;\;\nu}(p,\bar{n})$,
for each gluon, we can then associate one polarization tensor per gluon with
Eq.~\eqref{eq:coll_ggg_step1}. To simplify the corresponding result, we note that the
vertex factors in Eq.~\eqref{eq:scalar_current} and~\eqref{eq:decay_current} give
\begin{equation}\label{eq:currents_x_pols}
  \begin{split}
    S_\nu(p_i,p_j)d^{\mu\nu}(p_j,\bar{n})=\frac{2}{z_j}\,\frac{\tilde{p}_{i,j}^\mu}{p_{ij}^2}+\ldots\;,
    \quad\;\; D^\mu(p_i,p_j)=\frac{2\,\tilde{p}_{i,j}^\mu}{p_{ij}^2}+\ldots\;,\\
  \end{split}
\end{equation}
where the ellipses  represent terms proportional to $\bar{n}^\mu$ that vanish
upon multiplication with $d^\rho_{\;\mu}(p,\bar{n})$, independent of $p$.
We use these identities to simplify Eq.~\eqref{eq:coll_ggg_asym} as follows
\begin{equation}\label{eq:coll_ggg_asym2}
  \begin{split}
    \tilde{P}^{\rho\sigma,\alpha\beta}_{g\to g,\mu\nu,\rm(i)}(p_1,p_2)
    =&\;P^{\rho'\sigma',\alpha'\beta'}_{g\to g,\mu\nu,\rm(i)}(p_1,p_2)\,
    d^{\;\;\alpha}_{\alpha'}(p_2,\bar{n})d^{\;\;\beta}_{\beta'}(p_2,\bar{n})\,
    d^{\;\;\rho}_{\rho'}(p_1,\bar{n})d^{\;\;\sigma}_{\sigma'}(p_1,\bar{n})\\
    \approx&\;\frac{2C_A}{p_{12}^2}\,\bigg[\;
    \tilde{p}_{1,2}^\rho d_\mu^{\;\;\alpha}(p_{12},\bar{n})
    \bigg(\frac{1}{z_1z_2}\,\tilde{p}^\beta_{1,2}\,
    d_\nu^{\;\;\sigma}(p_{12},\bar{n})
    -\frac{1}{z_1}\,\tilde{p}_{1,2\,\nu}\,
    d^{\sigma\beta}(p_{12},\bar{n})\bigg)\\
    &\qquad\qquad\;
    -\frac{1}{z_2}\,\tilde{p}_{1,2}^\beta\tilde{p}_{1,2\,\mu}\,
    d_\nu^{\;\;\sigma}(p_{12},\bar{n})d^{\rho\alpha}(p_{12},\bar{n})\,\bigg]\;.
  \end{split}
\end{equation}
Let us now investigate the case where one of the final-state gluons
does not decay. This means in particular that the gluon can undergo subsequent
evolution, as long as it does not involve the current in Eq.~\eqref{eq:decay_current},
either in the amplitude, or its complex conjugate. Due to Eq.~\eqref{eq:axial_gauge}
being a projector in the strongly ordered limit, this corresponds to the trace
\begin{equation}\label{eq:coll_ggg_asym3}
  \begin{split}
    g_{\rho\sigma}\,\tilde{P}^{\rho\sigma,\alpha\beta}_{g\to g,\mu\nu,\rm(i)}(p_1,p_2)
    \approx&\;\frac{2C_A}{p_{12}^2}\,\frac{1}{z_2}\Big[\;
    \tilde{p}_{1,2\,\nu}\,d_\mu^{\;\;\alpha}(p_{12},\bar{n})
    -\tilde{p}_{1,2\,\mu}\,d_\nu^{\;\;\alpha}(p_{12},\bar{n})
    \,\Big]\,\tilde{p}_{1,2}^\beta\;.
  \end{split}
\end{equation}
We can use the fact that any physical gluon production tensor must be symmetric
in the Lorentz indices $\mu$ and $\nu$. This allows us to deduce that, in this case,
the interference contribution vanishes at leading power~\cite{Somogyi:2005xz}.
Because the triple-gluon vertex is crossing invariant, the same is also true if the
trace is taken over the Lorentz indices of the incoming gluon.
The only possibility for Eq.~\eqref{eq:coll_ggg_asym2} to contribute to the
matrix element in the strongly ordered soft or collinear limit is therefore
through a combination with $D^\mu$, either in the amplitude or its conjugate.
Note that both outgoing gluons must be combined with this tensor, otherwise
the result will again correspond to the trace in Eq.~\eqref{eq:coll_ggg_asym3}.
This requires an initial QCD dipole plus four or more gluons or quarks / antiquarks.
Equation~\eqref{eq:currents_x_pols} as well as Eqs.~\eqref{eq:g_to_sqsq_sd_split_coll}
and~\eqref{eq:zipjmzjpi_rad_coll} show that configurations which involve $D^\mu$
are not soft enhanced. Because both final-state gluons must undergo a decay
in order for Eq.~\eqref{eq:coll_ggg_asym2} to contribute, and because
Eq.~\eqref{eq:coll_ggg_asym2} itself has only an integrable soft singularity,
the interference effects are strongly suppressed overall.

We can make this explicit by investigating in more general terms the only configuration
under which the tensor in Eq.~\eqref{eq:coll_ggg_asym2} contributes to the squared
matrix element. As explained previously, any contraction of this tensor with a metric
tensor will yield a zero result. The only other symmetric tensor of rank two is of
type $M_\mu M_\nu$, with $M$ being an arbitrary four-vector. We thus find that the
only non-trivial contribution of Eq.~\eqref{eq:coll_ggg_asym2} to the squared
matrix element is through terms of the form
\begin{equation}\label{eq:coll_ggg_asym4}
  \begin{split}
    &K_\rho K_\sigma L_\alpha L_\beta M^\mu M^\nu\,
    \tilde{P}^{\rho\sigma,\alpha\beta}_{g\to g,\mu\nu,\rm(i)}(p_1,p_2)\\
    &\qquad\approx\frac{2C_A}{p_{12}^2}\,
    (\tilde{M}\tilde{L})(\tilde{M}\tilde{K})(\tilde{L}\tilde{K})\bigg[\;
    \frac{1}{z_1z_2}\frac{(\tilde{K}\tilde{p}_{1,2})(\tilde{p}_{1,2}\tilde{L})}{(\tilde{K}\tilde{L})}
    -\frac{1}{z_1}\frac{(\tilde{K}\tilde{p}_{1,2})(\tilde{p}_{1,2}\tilde{M})}{(\tilde{K}\tilde{M})}
    -\frac{1}{z_2}\frac{(\tilde{L}\tilde{p}_{1,2})(\tilde{p}_{1,2}\tilde{M})}{(\tilde{L}\tilde{M})}\,\bigg]\;,
  \end{split}
\end{equation}
where $\tilde{M}^\mu=d_\nu^{\;\;\mu}(p_{12},\bar{n})M^\nu$ is the component of $M^\mu$
transverse to the collinear direction $p_{12}$, and $\tilde{L}$ and $\tilde{K}$ are the
equivalently defined transverse components of $L$ and $K$.
To better understand the structure of Eq.~\eqref{eq:coll_ggg_asym4}, it is helpful to rewrite
the terms in the square bracket as functions of angles in the plane transverse to the jet axis.
This gives
\begin{equation}\label{eq:coll_ggg_asym5}
  \begin{split}
    &K_\rho K_\sigma L_\alpha L_\beta M^\mu M^\nu\,
    \tilde{P}^{\rho\sigma,\alpha\beta}_{g\to g,\mu\nu,\rm(i)}(p_1,p_2)\\
    &\qquad\approx 2C_A\,
    (\tilde{M}\tilde{L})(\tilde{M}\tilde{K})(\tilde{L}\tilde{K})\,z_2\bigg[\;
    \frac{\cos\theta_{\tilde{K}\tilde{p}_{1,2}}
    \cos\theta_{\tilde{L}\tilde{p}_{1,2}}}{\cos\theta_{\tilde{K}\tilde{L}}}
    -\frac{\cos\theta_{\tilde{K}\tilde{p}_{1,2}}
    \cos\theta_{\tilde{M}\tilde{p}_{1,2}}}{\cos\theta_{\tilde{K}\tilde{M}}}\,\bigg]
    +\bigg(\begin{array}{c}K\!\leftrightarrow \! L\\1\leftrightarrow 2\end{array}\bigg)\;.
  \end{split}
\end{equation}
Equation~\eqref{eq:coll_ggg_asym5} clearly shows that this contribution is not soft enhanced,
and that it vanishes upon integration over the azimuthal angle of $\tilde{p}_{1,2}$.
In order for it to contribute nontrivially to an observable, one must be able to resolve
four independent directions in the plane transverse to the jet axis, namely those of
$\tilde{K}$, $\tilde{L}$, $\tilde{M}$ and $\tilde{p}_{1,2}$. Because $\tilde{K}$, $\tilde{L}$
and $\tilde{M}$ are themselves defined through currents of the form of Eq.~\eqref{eq:currents_x_pols},
this means that any observable sensitive to Eq.~\eqref{eq:coll_ggg_asym5} must involve at least
five partons and must suppress soft contributions. Examples would be a five-point
energy correlator sensitive to all relative angles in the plane transverse to the jet axis,
or an energy flow polynomial of chromatic number five~\cite{Komiske:2017aww}.

Neglecting $\tilde{P}^{\rho\sigma,\alpha\beta}_{g\to g,\mu\nu,\rm(i)}$ in
Eq.~\eqref{eq:coll_ggg} can be seen as a variant of the factorization performed in
Ref.~\cite{Webber:1987uy}. This publication also stated that the non-factorizing components of the
gluon splitting function represent an acausal quantum correlation that cannot be accommodated
in a classical branching algorithm. However, Ref.~\cite{Webber:1987uy} is not based
on the same form of the gluon splitting that we propose in Eq.~\eqref{eq:coll_ggg_sym},
and their remainder function still includes a factorizable component. While this term is
a quantum correction and can therefore not be described by a local hidden variable theory,
it can still be simulated using Step~\ref{algo:decay} of our algorithm in Sec.~\ref{sec:algorithm}.
Using this improved proposal, Eq.~(6) of Ref.~\cite{Webber:1987uy} would vanish identically,
which further reduces the impact of the unaccounted for quantum effects on observables.

\bibliography{main}

\begin{thebibliography}{107}%
\makeatletter
\providecommand \@ifxundefined [1]{%
 \@ifx{#1\undefined}
}%
\providecommand \@ifnum [1]{%
 \ifnum #1\expandafter \@firstoftwo
 \else \expandafter \@secondoftwo
 \fi
}%
\providecommand \@ifx [1]{%
 \ifx #1\expandafter \@firstoftwo
 \else \expandafter \@secondoftwo
 \fi
}%
\providecommand \natexlab [1]{#1}%
\providecommand \enquote  [1]{``#1''}%
\providecommand \bibnamefont  [1]{#1}%
\providecommand \bibfnamefont [1]{#1}%
\providecommand \citenamefont [1]{#1}%
\providecommand \href@noop [0]{\@secondoftwo}%
\providecommand \href [0]{\begingroup \@sanitize@url \@href}%
\providecommand \@href[1]{\@@startlink{#1}\@@href}%
\providecommand \@@href[1]{\endgroup#1\@@endlink}%
\providecommand \@sanitize@url [0]{\catcode `\\12\catcode `\$12\catcode
  `\&12\catcode `\#12\catcode `\^12\catcode `\_12\catcode `\%12\relax}%
\providecommand \@@startlink[1]{}%
\providecommand \@@endlink[0]{}%
\providecommand \url  [0]{\begingroup\@sanitize@url \@url }%
\providecommand \@url [1]{\endgroup\@href {#1}{\urlprefix }}%
\providecommand \urlprefix  [0]{URL }%
\providecommand \Eprint [0]{\href }%
\providecommand \doibase [0]{http://dx.doi.org/}%
\providecommand \selectlanguage [0]{\@gobble}%
\providecommand \bibinfo  [0]{\@secondoftwo}%
\providecommand \bibfield  [0]{\@secondoftwo}%
\providecommand \translation [1]{[#1]}%
\providecommand \BibitemOpen [0]{}%
\providecommand \bibitemStop [0]{}%
\providecommand \bibitemNoStop [0]{.\EOS\space}%
\providecommand \EOS [0]{\spacefactor3000\relax}%
\providecommand \BibitemShut  [1]{\csname bibitem#1\endcsname}%
\let\auto@bib@innerbib\@empty
\bibitem [{\citenamefont {Webber}(1986)}]{Webber:1986mc}%
  \BibitemOpen
  \bibfield  {author} {\bibinfo {author} {\bibfnamefont {B.~R.}\ \bibnamefont
  {Webber}},\ }\href {\doibase 10.1146/annurev.ns.36.120186.001345} {\bibfield
  {journal} {\bibinfo  {journal} {Ann. Rev. Nucl. Part. Sci.}\ }\textbf
  {\bibinfo {volume} {36}},\ \bibinfo {pages} {253} (\bibinfo {year}
  {1986})}\BibitemShut {NoStop}%
\bibitem [{\citenamefont {Buckley}\ \emph {et~al.}(2011)\citenamefont {Buckley}
  \emph {et~al.}}]{Buckley:2011ms}%
  \BibitemOpen
  \bibfield  {author} {\bibinfo {author} {\bibfnamefont {A.}~\bibnamefont
  {Buckley}} \emph {et~al.},\ }\href {\doibase 10.1016/j.physrep.2011.03.005}
  {\bibfield  {journal} {\bibinfo  {journal} {Phys. Rept.}\ }\textbf {\bibinfo
  {volume} {504}},\ \bibinfo {pages} {145} (\bibinfo {year} {2011})},\ \Eprint
  {http://arxiv.org/abs/1101.2599} {arXiv:1101.2599 [hep-ph]} \BibitemShut
  {NoStop}%
\bibitem [{\citenamefont {Campbell}\ \emph {et~al.}(2024)\citenamefont
  {Campbell} \emph {et~al.}}]{Campbell:2022qmc}%
  \BibitemOpen
  \bibfield  {author} {\bibinfo {author} {\bibfnamefont {J.~M.}\ \bibnamefont
  {Campbell}} \emph {et~al.},\ }\href {\doibase 10.21468/SciPostPhys.16.5.130}
  {\bibfield  {journal} {\bibinfo  {journal} {SciPost Phys.}\ }\textbf
  {\bibinfo {volume} {16}},\ \bibinfo {pages} {130} (\bibinfo {year} {2024})},\
  \Eprint {http://arxiv.org/abs/2203.11110} {arXiv:2203.11110 [hep-ph]}
  \BibitemShut {NoStop}%
\bibitem [{\citenamefont {Amati}\ \emph {et~al.}(1980)\citenamefont {Amati},
  \citenamefont {Bassetto}, \citenamefont {Ciafaloni}, \citenamefont
  {Marchesini},\ and\ \citenamefont {Veneziano}}]{Amati:1980ch}%
  \BibitemOpen
  \bibfield  {author} {\bibinfo {author} {\bibfnamefont {D.}~\bibnamefont
  {Amati}}, \bibinfo {author} {\bibfnamefont {A.}~\bibnamefont {Bassetto}},
  \bibinfo {author} {\bibfnamefont {M.}~\bibnamefont {Ciafaloni}}, \bibinfo
  {author} {\bibfnamefont {G.}~\bibnamefont {Marchesini}}, \ and\ \bibinfo
  {author} {\bibfnamefont {G.}~\bibnamefont {Veneziano}},\ }\href {\doibase
  10.1016/0550-3213(80)90012-7} {\bibfield  {journal} {\bibinfo  {journal}
  {Nucl. Phys. B}\ }\textbf {\bibinfo {volume} {173}},\ \bibinfo {pages} {429}
  (\bibinfo {year} {1980})}\BibitemShut {NoStop}%
\bibitem [{\citenamefont {Catani}\ \emph {et~al.}(1991)\citenamefont {Catani},
  \citenamefont {Webber},\ and\ \citenamefont {Marchesini}}]{Catani:1990rr}%
  \BibitemOpen
  \bibfield  {author} {\bibinfo {author} {\bibfnamefont {S.}~\bibnamefont
  {Catani}}, \bibinfo {author} {\bibfnamefont {B.~R.}\ \bibnamefont {Webber}},
  \ and\ \bibinfo {author} {\bibfnamefont {G.}~\bibnamefont {Marchesini}},\
  }\href {\doibase 10.1016/0550-3213(91)90390-J} {\bibfield  {journal}
  {\bibinfo  {journal} {Nucl. Phys. B}\ }\textbf {\bibinfo {volume} {349}},\
  \bibinfo {pages} {635} (\bibinfo {year} {1991})}\BibitemShut {NoStop}%
\bibitem [{\citenamefont {Benedikt}\ \emph
  {et~al.}(2025{\natexlab{a}})\citenamefont {Benedikt} \emph
  {et~al.}}]{FCC:2025lpp}%
  \BibitemOpen
  \bibfield  {author} {\bibinfo {author} {\bibfnamefont {M.}~\bibnamefont
  {Benedikt}} \emph {et~al.} (\bibinfo {collaboration} {FCC}),\ }\href
  {\doibase 10.17181/CERN.9DKX.TDH9} {\  (\bibinfo {year}
  {2025}{\natexlab{a}}),\ 10.17181/CERN.9DKX.TDH9},\ \Eprint
  {http://arxiv.org/abs/2505.00272} {arXiv:2505.00272 [hep-ex]} \BibitemShut
  {NoStop}%
\bibitem [{\citenamefont {Benedikt}\ \emph
  {et~al.}(2025{\natexlab{b}})\citenamefont {Benedikt} \emph
  {et~al.}}]{FCC:2025uan}%
  \BibitemOpen
  \bibfield  {author} {\bibinfo {author} {\bibfnamefont {M.}~\bibnamefont
  {Benedikt}} \emph {et~al.} (\bibinfo {collaboration} {FCC}),\ }\href
  {\doibase 10.17181/CERN.EBAY.7W4X} {\  (\bibinfo {year}
  {2025}{\natexlab{b}}),\ 10.17181/CERN.EBAY.7W4X},\ \Eprint
  {http://arxiv.org/abs/2505.00274} {arXiv:2505.00274 [physics.acc-ph]}
  \BibitemShut {NoStop}%
\bibitem [{\citenamefont {Benedikt}\ \emph
  {et~al.}(2025{\natexlab{c}})\citenamefont {Benedikt} \emph
  {et~al.}}]{FCC:2025jtd}%
  \BibitemOpen
  \bibfield  {author} {\bibinfo {author} {\bibfnamefont {M.}~\bibnamefont
  {Benedikt}} \emph {et~al.} (\bibinfo {collaboration} {FCC}),\ }\href
  {\doibase 10.17181/CERN.I26X.V4VF} {\  (\bibinfo {year}
  {2025}{\natexlab{c}}),\ 10.17181/CERN.I26X.V4VF},\ \Eprint
  {http://arxiv.org/abs/2505.00273} {arXiv:2505.00273 [physics.acc-ph]}
  \BibitemShut {NoStop}%
\bibitem [{\citenamefont {Altheimer}\ \emph {et~al.}(2012)\citenamefont
  {Altheimer} \emph {et~al.}}]{Altheimer:2012mn}%
  \BibitemOpen
  \bibfield  {author} {\bibinfo {author} {\bibfnamefont {A.}~\bibnamefont
  {Altheimer}} \emph {et~al.},\ }\href {\doibase 10.1088/0954-3899/39/6/063001}
  {\bibfield  {journal} {\bibinfo  {journal} {J. Phys. G}\ }\textbf {\bibinfo
  {volume} {39}},\ \bibinfo {pages} {063001} (\bibinfo {year} {2012})},\
  \Eprint {http://arxiv.org/abs/1201.0008} {arXiv:1201.0008 [hep-ph]}
  \BibitemShut {NoStop}%
\bibitem [{\citenamefont {Larkoski}\ \emph {et~al.}(2020)\citenamefont
  {Larkoski}, \citenamefont {Moult},\ and\ \citenamefont
  {Nachman}}]{Larkoski:2017jix}%
  \BibitemOpen
  \bibfield  {author} {\bibinfo {author} {\bibfnamefont {A.~J.}\ \bibnamefont
  {Larkoski}}, \bibinfo {author} {\bibfnamefont {I.}~\bibnamefont {Moult}}, \
  and\ \bibinfo {author} {\bibfnamefont {B.}~\bibnamefont {Nachman}},\ }\href
  {\doibase 10.1016/j.physrep.2019.11.001} {\bibfield  {journal} {\bibinfo
  {journal} {Phys. Rept.}\ }\textbf {\bibinfo {volume} {841}},\ \bibinfo
  {pages} {1} (\bibinfo {year} {2020})},\ \Eprint
  {http://arxiv.org/abs/1709.04464} {arXiv:1709.04464 [hep-ph]} \BibitemShut
  {NoStop}%
\bibitem [{\citenamefont {Kogler}\ \emph {et~al.}(2019)\citenamefont {Kogler}
  \emph {et~al.}}]{Kogler:2018hem}%
  \BibitemOpen
  \bibfield  {author} {\bibinfo {author} {\bibfnamefont {R.}~\bibnamefont
  {Kogler}} \emph {et~al.},\ }\href {\doibase 10.1103/RevModPhys.91.045003}
  {\bibfield  {journal} {\bibinfo  {journal} {Rev. Mod. Phys.}\ }\textbf
  {\bibinfo {volume} {91}},\ \bibinfo {pages} {045003} (\bibinfo {year}
  {2019})},\ \Eprint {http://arxiv.org/abs/1803.06991} {arXiv:1803.06991
  [hep-ex]} \BibitemShut {NoStop}%
\bibitem [{\citenamefont {Bonilla}\ \emph {et~al.}(2022)\citenamefont {Bonilla}
  \emph {et~al.}}]{Bonilla:2022wzp}%
  \BibitemOpen
  \bibfield  {author} {\bibinfo {author} {\bibfnamefont {J.}~\bibnamefont
  {Bonilla}} \emph {et~al.},\ }\href {\doibase 10.3389/fphy.2022.897719}
  {\bibfield  {journal} {\bibinfo  {journal} {Front. in Phys.}\ }\textbf
  {\bibinfo {volume} {10}},\ \bibinfo {pages} {897719} (\bibinfo {year}
  {2022})},\ \Eprint {http://arxiv.org/abs/2203.07462} {arXiv:2203.07462
  [hep-ph]} \BibitemShut {NoStop}%
\bibitem [{\citenamefont {Webber}(1984)}]{Webber:1983if}%
  \BibitemOpen
  \bibfield  {author} {\bibinfo {author} {\bibfnamefont {B.~R.}\ \bibnamefont
  {Webber}},\ }\href {\doibase 10.1016/0550-3213(84)90333-X} {\bibfield
  {journal} {\bibinfo  {journal} {Nucl. Phys. B}\ }\textbf {\bibinfo {volume}
  {238}},\ \bibinfo {pages} {492} (\bibinfo {year} {1984})}\BibitemShut
  {NoStop}%
\bibitem [{\citenamefont {Bengtsson}\ \emph {et~al.}(1986)\citenamefont
  {Bengtsson}, \citenamefont {Sj{\"o}strand},\ and\ \citenamefont {van
  Zijl}}]{Bengtsson:1986gz}%
  \BibitemOpen
  \bibfield  {author} {\bibinfo {author} {\bibfnamefont {M.}~\bibnamefont
  {Bengtsson}}, \bibinfo {author} {\bibfnamefont {T.}~\bibnamefont
  {Sj{\"o}strand}}, \ and\ \bibinfo {author} {\bibfnamefont {M.}~\bibnamefont
  {van Zijl}},\ }\href {\doibase 10.1007/BF01441353} {\bibfield  {journal}
  {\bibinfo  {journal} {Z. Phys. C}\ }\textbf {\bibinfo {volume} {32}},\
  \bibinfo {pages} {67} (\bibinfo {year} {1986})}\BibitemShut {NoStop}%
\bibitem [{\citenamefont {Bengtsson}\ and\ \citenamefont
  {Sj{\"o}strand}(1987{\natexlab{a}})}]{Bengtsson:1986et}%
  \BibitemOpen
  \bibfield  {author} {\bibinfo {author} {\bibfnamefont {M.}~\bibnamefont
  {Bengtsson}}\ and\ \bibinfo {author} {\bibfnamefont {T.}~\bibnamefont
  {Sj{\"o}strand}},\ }\href {\doibase 10.1016/0550-3213(87)90407-X} {\bibfield
  {journal} {\bibinfo  {journal} {Nucl. Phys. B}\ }\textbf {\bibinfo {volume}
  {289}},\ \bibinfo {pages} {810} (\bibinfo {year}
  {1987}{\natexlab{a}})}\BibitemShut {NoStop}%
\bibitem [{\citenamefont {Marchesini}\ and\ \citenamefont
  {Webber}(1988)}]{Marchesini:1987cf}%
  \BibitemOpen
  \bibfield  {author} {\bibinfo {author} {\bibfnamefont {G.}~\bibnamefont
  {Marchesini}}\ and\ \bibinfo {author} {\bibfnamefont {B.~R.}\ \bibnamefont
  {Webber}},\ }\href {\doibase 10.1016/0550-3213(88)90089-2} {\bibfield
  {journal} {\bibinfo  {journal} {Nucl. Phys. B}\ }\textbf {\bibinfo {volume}
  {310}},\ \bibinfo {pages} {461} (\bibinfo {year} {1988})}\BibitemShut
  {NoStop}%
\bibitem [{\citenamefont {Andersson}\ \emph {et~al.}(1990)\citenamefont
  {Andersson}, \citenamefont {Gustafson},\ and\ \citenamefont
  {L{\"o}nnblad}}]{Andersson:1989ki}%
  \BibitemOpen
  \bibfield  {author} {\bibinfo {author} {\bibfnamefont {B.}~\bibnamefont
  {Andersson}}, \bibinfo {author} {\bibfnamefont {G.}~\bibnamefont
  {Gustafson}}, \ and\ \bibinfo {author} {\bibfnamefont {L.}~\bibnamefont
  {L{\"o}nnblad}},\ }\href {\doibase 10.1016/0550-3213(90)90355-H} {\bibfield
  {journal} {\bibinfo  {journal} {Nucl. Phys. B}\ }\textbf {\bibinfo {volume}
  {339}},\ \bibinfo {pages} {393} (\bibinfo {year} {1990})}\BibitemShut
  {NoStop}%
\bibitem [{\citenamefont {Bengtsson}\ and\ \citenamefont
  {Sj{\"o}strand}(1987{\natexlab{b}})}]{Bengtsson:1986hr}%
  \BibitemOpen
  \bibfield  {author} {\bibinfo {author} {\bibfnamefont {M.}~\bibnamefont
  {Bengtsson}}\ and\ \bibinfo {author} {\bibfnamefont {T.}~\bibnamefont
  {Sj{\"o}strand}},\ }\href {\doibase 10.1016/0370-2693(87)91031-8} {\bibfield
  {journal} {\bibinfo  {journal} {Phys. Lett. B}\ }\textbf {\bibinfo {volume}
  {185}},\ \bibinfo {pages} {435} (\bibinfo {year}
  {1987}{\natexlab{b}})}\BibitemShut {NoStop}%
\bibitem [{\citenamefont {Webber}(1987)}]{Webber:1987uy}%
  \BibitemOpen
  \bibfield  {author} {\bibinfo {author} {\bibfnamefont {B.~R.}\ \bibnamefont
  {Webber}},\ }\href {\doibase 10.1016/0370-2693(87)90461-8} {\bibfield
  {journal} {\bibinfo  {journal} {Phys. Lett. B}\ }\textbf {\bibinfo {volume}
  {193}},\ \bibinfo {pages} {91} (\bibinfo {year} {1987})}\BibitemShut
  {NoStop}%
\bibitem [{\citenamefont {Shatz}(1983)}]{Shatz:1983hv}%
  \BibitemOpen
  \bibfield  {author} {\bibinfo {author} {\bibfnamefont {M.~P.}\ \bibnamefont
  {Shatz}},\ }\href {\doibase 10.1016/0550-3213(83)90002-0} {\bibfield
  {journal} {\bibinfo  {journal} {Nucl. Phys. B}\ }\textbf {\bibinfo {volume}
  {224}},\ \bibinfo {pages} {218} (\bibinfo {year} {1983})}\BibitemShut
  {NoStop}%
\bibitem [{\citenamefont {Collins}(1988)}]{Collins:1987cp}%
  \BibitemOpen
  \bibfield  {author} {\bibinfo {author} {\bibfnamefont {J.~C.}\ \bibnamefont
  {Collins}},\ }\href {\doibase 10.1016/0550-3213(88)90654-2} {\bibfield
  {journal} {\bibinfo  {journal} {Nucl. Phys. B}\ }\textbf {\bibinfo {volume}
  {304}},\ \bibinfo {pages} {794} (\bibinfo {year} {1988})}\BibitemShut
  {NoStop}%
\bibitem [{\citenamefont {Knowles}(1988{\natexlab{a}})}]{Knowles:1987cu}%
  \BibitemOpen
  \bibfield  {author} {\bibinfo {author} {\bibfnamefont {I.~G.}\ \bibnamefont
  {Knowles}},\ }\href {\doibase 10.1016/0550-3213(88)90653-0} {\bibfield
  {journal} {\bibinfo  {journal} {Nucl. Phys. B}\ }\textbf {\bibinfo {volume}
  {304}},\ \bibinfo {pages} {767} (\bibinfo {year}
  {1988}{\natexlab{a}})}\BibitemShut {NoStop}%
\bibitem [{\citenamefont {Knowles}(1988{\natexlab{b}})}]{Knowles:1988vs}%
  \BibitemOpen
  \bibfield  {author} {\bibinfo {author} {\bibfnamefont {I.~G.}\ \bibnamefont
  {Knowles}},\ }\href {\doibase 10.1016/0550-3213(88)90092-2} {\bibfield
  {journal} {\bibinfo  {journal} {Nucl. Phys. B}\ }\textbf {\bibinfo {volume}
  {310}},\ \bibinfo {pages} {571} (\bibinfo {year}
  {1988}{\natexlab{b}})}\BibitemShut {NoStop}%
\bibitem [{\citenamefont {Knowles}(1990)}]{Knowles:1988hu}%
  \BibitemOpen
  \bibfield  {author} {\bibinfo {author} {\bibfnamefont {I.~G.}\ \bibnamefont
  {Knowles}},\ }\href {\doibase 10.1016/0010-4655(90)90063-7} {\bibfield
  {journal} {\bibinfo  {journal} {Comput. Phys. Commun.}\ }\textbf {\bibinfo
  {volume} {58}},\ \bibinfo {pages} {271} (\bibinfo {year} {1990})}\BibitemShut
  {NoStop}%
\bibitem [{\citenamefont {van Beekveld}\ \emph {et~al.}(2022)\citenamefont {van
  Beekveld}, \citenamefont {Ferrario~Ravasio}, \citenamefont {Hamilton},
  \citenamefont {Salam}, \citenamefont {Soto-Ontoso}, \citenamefont {Soyez},\
  and\ \citenamefont {Verheyen}}]{vanBeekveld:2022ukn}%
  \BibitemOpen
  \bibfield  {author} {\bibinfo {author} {\bibfnamefont {M.}~\bibnamefont {van
  Beekveld}}, \bibinfo {author} {\bibfnamefont {S.}~\bibnamefont
  {Ferrario~Ravasio}}, \bibinfo {author} {\bibfnamefont {K.}~\bibnamefont
  {Hamilton}}, \bibinfo {author} {\bibfnamefont {G.~P.}\ \bibnamefont {Salam}},
  \bibinfo {author} {\bibfnamefont {A.}~\bibnamefont {Soto-Ontoso}}, \bibinfo
  {author} {\bibfnamefont {G.}~\bibnamefont {Soyez}}, \ and\ \bibinfo {author}
  {\bibfnamefont {R.}~\bibnamefont {Verheyen}},\ }\href {\doibase
  10.1007/JHEP11(2022)020} {\bibfield  {journal} {\bibinfo  {journal} {JHEP}\
  }\textbf {\bibinfo {volume} {11}},\ \bibinfo {pages} {020} (\bibinfo {year}
  {2022})},\ \Eprint {http://arxiv.org/abs/2207.09467} {arXiv:2207.09467
  [hep-ph]} \BibitemShut {NoStop}%
\bibitem [{\citenamefont {Nagy}\ and\ \citenamefont
  {Soper}(2005)}]{Nagy:2005aa}%
  \BibitemOpen
  \bibfield  {author} {\bibinfo {author} {\bibfnamefont {Z.}~\bibnamefont
  {Nagy}}\ and\ \bibinfo {author} {\bibfnamefont {D.~E.}\ \bibnamefont
  {Soper}},\ }\href {\doibase 10.1088/1126-6708/2005/10/024} {\bibfield
  {journal} {\bibinfo  {journal} {JHEP}\ }\textbf {\bibinfo {volume} {10}},\
  \bibinfo {pages} {024} (\bibinfo {year} {2005})},\ \Eprint
  {http://arxiv.org/abs/hep-ph/0503053} {arXiv:hep-ph/0503053} \BibitemShut
  {NoStop}%
\bibitem [{\citenamefont {Nagy}\ and\ \citenamefont
  {Soper}(2006)}]{Nagy:2006kb}%
  \BibitemOpen
  \bibfield  {author} {\bibinfo {author} {\bibfnamefont {Z.}~\bibnamefont
  {Nagy}}\ and\ \bibinfo {author} {\bibfnamefont {D.~E.}\ \bibnamefont
  {Soper}},\ }in\ \href {\doibase 10.1142/9789812773524_0010} {\emph {\bibinfo
  {booktitle} {{Ringberg Workshop on New Trends in HERA Physics 2005}}}}\
  (\bibinfo {year} {2006})\ pp.\ \bibinfo {pages} {101--123},\ \Eprint
  {http://arxiv.org/abs/hep-ph/0601021} {arXiv:hep-ph/0601021} \BibitemShut
  {NoStop}%
\bibitem [{\citenamefont {Schumann}\ and\ \citenamefont
  {Krauss}(2008)}]{Schumann:2007mg}%
  \BibitemOpen
  \bibfield  {author} {\bibinfo {author} {\bibfnamefont {S.}~\bibnamefont
  {Schumann}}\ and\ \bibinfo {author} {\bibfnamefont {F.}~\bibnamefont
  {Krauss}},\ }\href {\doibase 10.1088/1126-6708/2008/03/038} {\bibfield
  {journal} {\bibinfo  {journal} {JHEP}\ }\textbf {\bibinfo {volume} {03}},\
  \bibinfo {pages} {038} (\bibinfo {year} {2008})},\ \Eprint
  {http://arxiv.org/abs/0709.1027} {arXiv:0709.1027 [hep-ph]} \BibitemShut
  {NoStop}%
\bibitem [{\citenamefont {Giele}\ \emph {et~al.}(2008)\citenamefont {Giele},
  \citenamefont {Kosower},\ and\ \citenamefont {Skands}}]{Giele:2007di}%
  \BibitemOpen
  \bibfield  {author} {\bibinfo {author} {\bibfnamefont {W.~T.}\ \bibnamefont
  {Giele}}, \bibinfo {author} {\bibfnamefont {D.~A.}\ \bibnamefont {Kosower}},
  \ and\ \bibinfo {author} {\bibfnamefont {P.~Z.}\ \bibnamefont {Skands}},\
  }\href {\doibase 10.1103/PhysRevD.78.014026} {\bibfield  {journal} {\bibinfo
  {journal} {Phys. Rev. D}\ }\textbf {\bibinfo {volume} {78}},\ \bibinfo
  {pages} {014026} (\bibinfo {year} {2008})},\ \Eprint
  {http://arxiv.org/abs/0707.3652} {arXiv:0707.3652 [hep-ph]} \BibitemShut
  {NoStop}%
\bibitem [{\citenamefont {Pl{\"a}tzer}\ and\ \citenamefont
  {Gieseke}(2011)}]{Platzer:2009jq}%
  \BibitemOpen
  \bibfield  {author} {\bibinfo {author} {\bibfnamefont {S.}~\bibnamefont
  {Pl{\"a}tzer}}\ and\ \bibinfo {author} {\bibfnamefont {S.}~\bibnamefont
  {Gieseke}},\ }\href {\doibase 10.1007/JHEP01(2011)024} {\bibfield  {journal}
  {\bibinfo  {journal} {JHEP}\ }\textbf {\bibinfo {volume} {01}},\ \bibinfo
  {pages} {024} (\bibinfo {year} {2011})},\ \Eprint
  {http://arxiv.org/abs/0909.5593} {arXiv:0909.5593 [hep-ph]} \BibitemShut
  {NoStop}%
\bibitem [{\citenamefont {H{\"o}che}\ and\ \citenamefont
  {Prestel}(2015)}]{Hoche:2015sya}%
  \BibitemOpen
  \bibfield  {author} {\bibinfo {author} {\bibfnamefont {S.}~\bibnamefont
  {H{\"o}che}}\ and\ \bibinfo {author} {\bibfnamefont {S.}~\bibnamefont
  {Prestel}},\ }\href {\doibase 10.1140/epjc/s10052-015-3684-2} {\bibfield
  {journal} {\bibinfo  {journal} {Eur. Phys. J. C}\ }\textbf {\bibinfo {volume}
  {75}},\ \bibinfo {pages} {461} (\bibinfo {year} {2015})},\ \Eprint
  {http://arxiv.org/abs/1506.05057} {arXiv:1506.05057 [hep-ph]} \BibitemShut
  {NoStop}%
\bibitem [{\citenamefont {Fischer}\ \emph {et~al.}(2016)\citenamefont
  {Fischer}, \citenamefont {Prestel}, \citenamefont {Ritzmann},\ and\
  \citenamefont {Skands}}]{Fischer:2016vfv}%
  \BibitemOpen
  \bibfield  {author} {\bibinfo {author} {\bibfnamefont {N.}~\bibnamefont
  {Fischer}}, \bibinfo {author} {\bibfnamefont {S.}~\bibnamefont {Prestel}},
  \bibinfo {author} {\bibfnamefont {M.}~\bibnamefont {Ritzmann}}, \ and\
  \bibinfo {author} {\bibfnamefont {P.}~\bibnamefont {Skands}},\ }\href
  {\doibase 10.1140/epjc/s10052-016-4429-6} {\bibfield  {journal} {\bibinfo
  {journal} {Eur. Phys. J. C}\ }\textbf {\bibinfo {volume} {76}},\ \bibinfo
  {pages} {589} (\bibinfo {year} {2016})},\ \Eprint
  {http://arxiv.org/abs/1605.06142} {arXiv:1605.06142 [hep-ph]} \BibitemShut
  {NoStop}%
\bibitem [{\citenamefont {Cabouat}\ and\ \citenamefont
  {Sj{\"o}strand}(2018)}]{Cabouat:2017rzi}%
  \BibitemOpen
  \bibfield  {author} {\bibinfo {author} {\bibfnamefont {B.}~\bibnamefont
  {Cabouat}}\ and\ \bibinfo {author} {\bibfnamefont {T.}~\bibnamefont
  {Sj{\"o}strand}},\ }\href {\doibase 10.1140/epjc/s10052-018-5645-z}
  {\bibfield  {journal} {\bibinfo  {journal} {Eur. Phys. J. C}\ }\textbf
  {\bibinfo {volume} {78}},\ \bibinfo {pages} {226} (\bibinfo {year} {2018})},\
  \Eprint {http://arxiv.org/abs/1710.00391} {arXiv:1710.00391 [hep-ph]}
  \BibitemShut {NoStop}%
\bibitem [{\citenamefont {Nagy}\ and\ \citenamefont
  {Soper}(2012)}]{Nagy:2012bt}%
  \BibitemOpen
  \bibfield  {author} {\bibinfo {author} {\bibfnamefont {Z.}~\bibnamefont
  {Nagy}}\ and\ \bibinfo {author} {\bibfnamefont {D.~E.}\ \bibnamefont
  {Soper}},\ }\href {\doibase 10.1007/JHEP06(2012)044} {\bibfield  {journal}
  {\bibinfo  {journal} {JHEP}\ }\textbf {\bibinfo {volume} {06}},\ \bibinfo
  {pages} {044} (\bibinfo {year} {2012})},\ \Eprint
  {http://arxiv.org/abs/1202.4496} {arXiv:1202.4496 [hep-ph]} \BibitemShut
  {NoStop}%
\bibitem [{\citenamefont {Pl{\"a}tzer}\ and\ \citenamefont
  {Sj{\"o}dahl}(2012)}]{Platzer:2012np}%
  \BibitemOpen
  \bibfield  {author} {\bibinfo {author} {\bibfnamefont {S.}~\bibnamefont
  {Pl{\"a}tzer}}\ and\ \bibinfo {author} {\bibfnamefont {M.}~\bibnamefont
  {Sj{\"o}dahl}},\ }\href {\doibase 10.1007/JHEP07(2012)042} {\bibfield
  {journal} {\bibinfo  {journal} {JHEP}\ }\textbf {\bibinfo {volume} {07}},\
  \bibinfo {pages} {042} (\bibinfo {year} {2012})},\ \Eprint
  {http://arxiv.org/abs/1201.0260} {arXiv:1201.0260 [hep-ph]} \BibitemShut
  {NoStop}%
\bibitem [{\citenamefont {Nagy}\ and\ \citenamefont
  {Soper}(2014)}]{Nagy:2014mqa}%
  \BibitemOpen
  \bibfield  {author} {\bibinfo {author} {\bibfnamefont {Z.}~\bibnamefont
  {Nagy}}\ and\ \bibinfo {author} {\bibfnamefont {D.~E.}\ \bibnamefont
  {Soper}},\ }\href {\doibase 10.1007/JHEP06(2014)097} {\bibfield  {journal}
  {\bibinfo  {journal} {JHEP}\ }\textbf {\bibinfo {volume} {06}},\ \bibinfo
  {pages} {097} (\bibinfo {year} {2014})},\ \Eprint
  {http://arxiv.org/abs/1401.6364} {arXiv:1401.6364 [hep-ph]} \BibitemShut
  {NoStop}%
\bibitem [{\citenamefont {Nagy}\ and\ \citenamefont
  {Soper}(2015)}]{Nagy:2015hwa}%
  \BibitemOpen
  \bibfield  {author} {\bibinfo {author} {\bibfnamefont {Z.}~\bibnamefont
  {Nagy}}\ and\ \bibinfo {author} {\bibfnamefont {D.~E.}\ \bibnamefont
  {Soper}},\ }\href {\doibase 10.1007/JHEP07(2015)119} {\bibfield  {journal}
  {\bibinfo  {journal} {JHEP}\ }\textbf {\bibinfo {volume} {07}},\ \bibinfo
  {pages} {119} (\bibinfo {year} {2015})},\ \Eprint
  {http://arxiv.org/abs/1501.00778} {arXiv:1501.00778 [hep-ph]} \BibitemShut
  {NoStop}%
\bibitem [{\citenamefont {Pl{\"a}tzer}\ \emph {et~al.}(2018)\citenamefont
  {Pl{\"a}tzer}, \citenamefont {Sj{\"o}dahl},\ and\ \citenamefont
  {Thor{\'e}n}}]{Platzer:2018pmd}%
  \BibitemOpen
  \bibfield  {author} {\bibinfo {author} {\bibfnamefont {S.}~\bibnamefont
  {Pl{\"a}tzer}}, \bibinfo {author} {\bibfnamefont {M.}~\bibnamefont
  {Sj{\"o}dahl}}, \ and\ \bibinfo {author} {\bibfnamefont {J.}~\bibnamefont
  {Thor{\'e}n}},\ }\href {\doibase 10.1007/JHEP11(2018)009} {\bibfield
  {journal} {\bibinfo  {journal} {JHEP}\ }\textbf {\bibinfo {volume} {11}},\
  \bibinfo {pages} {009} (\bibinfo {year} {2018})},\ \Eprint
  {http://arxiv.org/abs/1808.00332} {arXiv:1808.00332 [hep-ph]} \BibitemShut
  {NoStop}%
\bibitem [{\citenamefont {Isaacson}\ and\ \citenamefont
  {Prestel}(2019)}]{Isaacson:2018zdi}%
  \BibitemOpen
  \bibfield  {author} {\bibinfo {author} {\bibfnamefont {J.}~\bibnamefont
  {Isaacson}}\ and\ \bibinfo {author} {\bibfnamefont {S.}~\bibnamefont
  {Prestel}},\ }\href {\doibase 10.1103/PhysRevD.99.014021} {\bibfield
  {journal} {\bibinfo  {journal} {Phys. Rev. D}\ }\textbf {\bibinfo {volume}
  {99}},\ \bibinfo {pages} {014021} (\bibinfo {year} {2019})},\ \Eprint
  {http://arxiv.org/abs/1806.10102} {arXiv:1806.10102 [hep-ph]} \BibitemShut
  {NoStop}%
\bibitem [{\citenamefont {Nagy}\ and\ \citenamefont
  {Soper}(2019{\natexlab{a}})}]{Nagy:2019rwb}%
  \BibitemOpen
  \bibfield  {author} {\bibinfo {author} {\bibfnamefont {Z.}~\bibnamefont
  {Nagy}}\ and\ \bibinfo {author} {\bibfnamefont {D.~E.}\ \bibnamefont
  {Soper}},\ }\href {\doibase 10.1103/PhysRevD.100.074005} {\bibfield
  {journal} {\bibinfo  {journal} {Phys. Rev. D}\ }\textbf {\bibinfo {volume}
  {100}},\ \bibinfo {pages} {074005} (\bibinfo {year} {2019}{\natexlab{a}})},\
  \Eprint {http://arxiv.org/abs/1908.11420} {arXiv:1908.11420 [hep-ph]}
  \BibitemShut {NoStop}%
\bibitem [{\citenamefont {Nagy}\ and\ \citenamefont
  {Soper}(2019{\natexlab{b}})}]{Nagy:2019pjp}%
  \BibitemOpen
  \bibfield  {author} {\bibinfo {author} {\bibfnamefont {Z.}~\bibnamefont
  {Nagy}}\ and\ \bibinfo {author} {\bibfnamefont {D.~E.}\ \bibnamefont
  {Soper}},\ }\href {\doibase 10.1103/PhysRevD.99.054009} {\bibfield  {journal}
  {\bibinfo  {journal} {Phys. Rev. D}\ }\textbf {\bibinfo {volume} {99}},\
  \bibinfo {pages} {054009} (\bibinfo {year} {2019}{\natexlab{b}})},\ \Eprint
  {http://arxiv.org/abs/1902.02105} {arXiv:1902.02105 [hep-ph]} \BibitemShut
  {NoStop}%
\bibitem [{\citenamefont {Forshaw}\ \emph {et~al.}(2019)\citenamefont
  {Forshaw}, \citenamefont {Holguin},\ and\ \citenamefont
  {Pl{\"a}tzer}}]{Forshaw:2019ver}%
  \BibitemOpen
  \bibfield  {author} {\bibinfo {author} {\bibfnamefont {J.~R.}\ \bibnamefont
  {Forshaw}}, \bibinfo {author} {\bibfnamefont {J.}~\bibnamefont {Holguin}}, \
  and\ \bibinfo {author} {\bibfnamefont {S.}~\bibnamefont {Pl{\"a}tzer}},\
  }\href {\doibase 10.1007/JHEP08(2019)145} {\bibfield  {journal} {\bibinfo
  {journal} {JHEP}\ }\textbf {\bibinfo {volume} {08}},\ \bibinfo {pages} {145}
  (\bibinfo {year} {2019})},\ \Eprint {http://arxiv.org/abs/1905.08686}
  {arXiv:1905.08686 [hep-ph]} \BibitemShut {NoStop}%
\bibitem [{\citenamefont {H{\"o}che}\ and\ \citenamefont
  {Reichelt}(2021)}]{Hoche:2020pxj}%
  \BibitemOpen
  \bibfield  {author} {\bibinfo {author} {\bibfnamefont {S.}~\bibnamefont
  {H{\"o}che}}\ and\ \bibinfo {author} {\bibfnamefont {D.}~\bibnamefont
  {Reichelt}},\ }\href {\doibase 10.1103/PhysRevD.104.034006} {\bibfield
  {journal} {\bibinfo  {journal} {Phys. Rev. D}\ }\textbf {\bibinfo {volume}
  {104}},\ \bibinfo {pages} {034006} (\bibinfo {year} {2021})},\ \Eprint
  {http://arxiv.org/abs/2001.11492} {arXiv:2001.11492 [hep-ph]} \BibitemShut
  {NoStop}%
\bibitem [{\citenamefont {De~Angelis}\ \emph {et~al.}(2021)\citenamefont
  {De~Angelis}, \citenamefont {Forshaw},\ and\ \citenamefont
  {Pl{\"a}tzer}}]{DeAngelis:2020rvq}%
  \BibitemOpen
  \bibfield  {author} {\bibinfo {author} {\bibfnamefont {M.}~\bibnamefont
  {De~Angelis}}, \bibinfo {author} {\bibfnamefont {J.~R.}\ \bibnamefont
  {Forshaw}}, \ and\ \bibinfo {author} {\bibfnamefont {S.}~\bibnamefont
  {Pl{\"a}tzer}},\ }\href {\doibase 10.1103/PhysRevLett.126.112001} {\bibfield
  {journal} {\bibinfo  {journal} {Phys. Rev. Lett.}\ }\textbf {\bibinfo
  {volume} {126}},\ \bibinfo {pages} {112001} (\bibinfo {year} {2021})},\
  \Eprint {http://arxiv.org/abs/2007.09648} {arXiv:2007.09648 [hep-ph]}
  \BibitemShut {NoStop}%
\bibitem [{\citenamefont {Holguin}\ \emph {et~al.}(2021)\citenamefont
  {Holguin}, \citenamefont {Forshaw},\ and\ \citenamefont
  {Pl{\"a}tzer}}]{Holguin:2020joq}%
  \BibitemOpen
  \bibfield  {author} {\bibinfo {author} {\bibfnamefont {J.}~\bibnamefont
  {Holguin}}, \bibinfo {author} {\bibfnamefont {J.~R.}\ \bibnamefont
  {Forshaw}}, \ and\ \bibinfo {author} {\bibfnamefont {S.}~\bibnamefont
  {Pl{\"a}tzer}},\ }\href {\doibase 10.1140/epjc/s10052-021-09145-1} {\bibfield
   {journal} {\bibinfo  {journal} {Eur. Phys. J. C}\ }\textbf {\bibinfo
  {volume} {81}},\ \bibinfo {pages} {364} (\bibinfo {year} {2021})},\ \Eprint
  {http://arxiv.org/abs/2011.15087} {arXiv:2011.15087 [hep-ph]} \BibitemShut
  {NoStop}%
\bibitem [{\citenamefont {Gustafson}(1993)}]{Gustafson:1992uh}%
  \BibitemOpen
  \bibfield  {author} {\bibinfo {author} {\bibfnamefont {G.}~\bibnamefont
  {Gustafson}},\ }\href {\doibase 10.1016/0550-3213(93)90203-2} {\bibfield
  {journal} {\bibinfo  {journal} {Nucl. Phys. B}\ }\textbf {\bibinfo {volume}
  {392}},\ \bibinfo {pages} {251} (\bibinfo {year} {1993})}\BibitemShut
  {NoStop}%
\bibitem [{\citenamefont {Hamilton}\ \emph {et~al.}(2021)\citenamefont
  {Hamilton}, \citenamefont {Medves}, \citenamefont {Salam}, \citenamefont
  {Scyboz},\ and\ \citenamefont {Soyez}}]{Hamilton:2020rcu}%
  \BibitemOpen
  \bibfield  {author} {\bibinfo {author} {\bibfnamefont {K.}~\bibnamefont
  {Hamilton}}, \bibinfo {author} {\bibfnamefont {R.}~\bibnamefont {Medves}},
  \bibinfo {author} {\bibfnamefont {G.~P.}\ \bibnamefont {Salam}}, \bibinfo
  {author} {\bibfnamefont {L.}~\bibnamefont {Scyboz}}, \ and\ \bibinfo {author}
  {\bibfnamefont {G.}~\bibnamefont {Soyez}},\ }\href {\doibase
  10.1007/JHEP03(2021)041} {\bibfield  {journal} {\bibinfo  {journal} {JHEP}\
  }\textbf {\bibinfo {volume} {03}},\ \bibinfo {pages} {041} (\bibinfo {year}
  {2021})},\ \Eprint {http://arxiv.org/abs/2011.10054} {arXiv:2011.10054
  [hep-ph]} \BibitemShut {NoStop}%
\bibitem [{\citenamefont {Dasgupta}\ \emph {et~al.}(2018)\citenamefont
  {Dasgupta}, \citenamefont {Dreyer}, \citenamefont {Hamilton}, \citenamefont
  {Monni},\ and\ \citenamefont {Salam}}]{Dasgupta:2018nvj}%
  \BibitemOpen
  \bibfield  {author} {\bibinfo {author} {\bibfnamefont {M.}~\bibnamefont
  {Dasgupta}}, \bibinfo {author} {\bibfnamefont {F.~A.}\ \bibnamefont
  {Dreyer}}, \bibinfo {author} {\bibfnamefont {K.}~\bibnamefont {Hamilton}},
  \bibinfo {author} {\bibfnamefont {P.~F.}\ \bibnamefont {Monni}}, \ and\
  \bibinfo {author} {\bibfnamefont {G.~P.}\ \bibnamefont {Salam}},\ }\href
  {\doibase 10.1007/JHEP09(2018)033} {\bibfield  {journal} {\bibinfo  {journal}
  {JHEP}\ }\textbf {\bibinfo {volume} {09}},\ \bibinfo {pages} {033} (\bibinfo
  {year} {2018})},\ \bibinfo {note} {[Erratum: JHEP 03, 083 (2020)]},\ \Eprint
  {http://arxiv.org/abs/1805.09327} {arXiv:1805.09327 [hep-ph]} \BibitemShut
  {NoStop}%
\bibitem [{\citenamefont {Dasgupta}\ \emph {et~al.}(2020)\citenamefont
  {Dasgupta}, \citenamefont {Dreyer}, \citenamefont {Hamilton}, \citenamefont
  {Monni}, \citenamefont {Salam},\ and\ \citenamefont
  {Soyez}}]{Dasgupta:2020fwr}%
  \BibitemOpen
  \bibfield  {author} {\bibinfo {author} {\bibfnamefont {M.}~\bibnamefont
  {Dasgupta}}, \bibinfo {author} {\bibfnamefont {F.~A.}\ \bibnamefont
  {Dreyer}}, \bibinfo {author} {\bibfnamefont {K.}~\bibnamefont {Hamilton}},
  \bibinfo {author} {\bibfnamefont {P.~F.}\ \bibnamefont {Monni}}, \bibinfo
  {author} {\bibfnamefont {G.~P.}\ \bibnamefont {Salam}}, \ and\ \bibinfo
  {author} {\bibfnamefont {G.}~\bibnamefont {Soyez}},\ }\href {\doibase
  10.1103/PhysRevLett.125.052002} {\bibfield  {journal} {\bibinfo  {journal}
  {Phys. Rev. Lett.}\ }\textbf {\bibinfo {volume} {125}},\ \bibinfo {pages}
  {052002} (\bibinfo {year} {2020})},\ \Eprint
  {http://arxiv.org/abs/2002.11114} {arXiv:2002.11114 [hep-ph]} \BibitemShut
  {NoStop}%
\bibitem [{\citenamefont {Herren}\ \emph {et~al.}(2023)\citenamefont {Herren},
  \citenamefont {H{\"o}che}, \citenamefont {Krauss}, \citenamefont {Reichelt},\
  and\ \citenamefont {Sch{\"o}nherr}}]{Herren:2022jej}%
  \BibitemOpen
  \bibfield  {author} {\bibinfo {author} {\bibfnamefont {F.}~\bibnamefont
  {Herren}}, \bibinfo {author} {\bibfnamefont {S.}~\bibnamefont {H{\"o}che}},
  \bibinfo {author} {\bibfnamefont {F.}~\bibnamefont {Krauss}}, \bibinfo
  {author} {\bibfnamefont {D.}~\bibnamefont {Reichelt}}, \ and\ \bibinfo
  {author} {\bibfnamefont {M.}~\bibnamefont {Sch{\"o}nherr}},\ }\href {\doibase
  10.1007/JHEP10(2023)091} {\bibfield  {journal} {\bibinfo  {journal} {JHEP}\
  }\textbf {\bibinfo {volume} {10}},\ \bibinfo {pages} {091} (\bibinfo {year}
  {2023})},\ \Eprint {http://arxiv.org/abs/2208.06057} {arXiv:2208.06057
  [hep-ph]} \BibitemShut {NoStop}%
\bibitem [{\citenamefont {Assi}\ and\ \citenamefont
  {H{\"o}che}(2024)}]{Assi:2023rbu}%
  \BibitemOpen
  \bibfield  {author} {\bibinfo {author} {\bibfnamefont {B.}~\bibnamefont
  {Assi}}\ and\ \bibinfo {author} {\bibfnamefont {S.}~\bibnamefont
  {H{\"o}che}},\ }\href {\doibase 10.1103/PhysRevD.109.114008} {\bibfield
  {journal} {\bibinfo  {journal} {Phys. Rev. D}\ }\textbf {\bibinfo {volume}
  {109}},\ \bibinfo {pages} {114008} (\bibinfo {year} {2024})},\ \Eprint
  {http://arxiv.org/abs/2307.00728} {arXiv:2307.00728 [hep-ph]} \BibitemShut
  {NoStop}%
\bibitem [{\citenamefont {H{\"o}che}\ \emph
  {et~al.}(2025{\natexlab{a}})\citenamefont {H{\"o}che}, \citenamefont
  {Krauss},\ and\ \citenamefont {Reichelt}}]{Hoche:2024dee}%
  \BibitemOpen
  \bibfield  {author} {\bibinfo {author} {\bibfnamefont {S.}~\bibnamefont
  {H{\"o}che}}, \bibinfo {author} {\bibfnamefont {F.}~\bibnamefont {Krauss}}, \
  and\ \bibinfo {author} {\bibfnamefont {D.}~\bibnamefont {Reichelt}},\ }\href
  {\doibase 10.1103/PhysRevD.111.094032} {\bibfield  {journal} {\bibinfo
  {journal} {Phys. Rev. D}\ }\textbf {\bibinfo {volume} {111}},\ \bibinfo
  {pages} {094032} (\bibinfo {year} {2025}{\natexlab{a}})},\ \Eprint
  {http://arxiv.org/abs/2404.14360} {arXiv:2404.14360 [hep-ph]} \BibitemShut
  {NoStop}%
\bibitem [{\citenamefont {Preuss}(2024)}]{Preuss:2024vyu}%
  \BibitemOpen
  \bibfield  {author} {\bibinfo {author} {\bibfnamefont {C.~T.}\ \bibnamefont
  {Preuss}},\ }\href {\doibase 10.1007/JHEP07(2024)161} {\bibfield  {journal}
  {\bibinfo  {journal} {JHEP}\ }\textbf {\bibinfo {volume} {07}},\ \bibinfo
  {pages} {161} (\bibinfo {year} {2024})},\ \Eprint
  {http://arxiv.org/abs/2403.19452} {arXiv:2403.19452 [hep-ph]} \BibitemShut
  {NoStop}%
\bibitem [{\citenamefont {H{\"o}che}\ and\ \citenamefont
  {Prestel}(2017)}]{Hoche:2017iem}%
  \BibitemOpen
  \bibfield  {author} {\bibinfo {author} {\bibfnamefont {S.}~\bibnamefont
  {H{\"o}che}}\ and\ \bibinfo {author} {\bibfnamefont {S.}~\bibnamefont
  {Prestel}},\ }\href {\doibase 10.1103/PhysRevD.96.074017} {\bibfield
  {journal} {\bibinfo  {journal} {Phys. Rev. D}\ }\textbf {\bibinfo {volume}
  {96}},\ \bibinfo {pages} {074017} (\bibinfo {year} {2017})},\ \Eprint
  {http://arxiv.org/abs/1705.00742} {arXiv:1705.00742 [hep-ph]} \BibitemShut
  {NoStop}%
\bibitem [{\citenamefont {Dulat}\ \emph {et~al.}(2018)\citenamefont {Dulat},
  \citenamefont {H{\"o}che},\ and\ \citenamefont {Prestel}}]{Dulat:2018vuy}%
  \BibitemOpen
  \bibfield  {author} {\bibinfo {author} {\bibfnamefont {F.}~\bibnamefont
  {Dulat}}, \bibinfo {author} {\bibfnamefont {S.}~\bibnamefont {H{\"o}che}}, \
  and\ \bibinfo {author} {\bibfnamefont {S.}~\bibnamefont {Prestel}},\ }\href
  {\doibase 10.1103/PhysRevD.98.074013} {\bibfield  {journal} {\bibinfo
  {journal} {Phys. Rev. D}\ }\textbf {\bibinfo {volume} {98}},\ \bibinfo
  {pages} {074013} (\bibinfo {year} {2018})},\ \Eprint
  {http://arxiv.org/abs/1805.03757} {arXiv:1805.03757 [hep-ph]} \BibitemShut
  {NoStop}%
\bibitem [{\citenamefont {Gellersen}\ \emph {et~al.}(2022)\citenamefont
  {Gellersen}, \citenamefont {H{\"o}che},\ and\ \citenamefont
  {Prestel}}]{Gellersen:2021eci}%
  \BibitemOpen
  \bibfield  {author} {\bibinfo {author} {\bibfnamefont {L.}~\bibnamefont
  {Gellersen}}, \bibinfo {author} {\bibfnamefont {S.}~\bibnamefont
  {H{\"o}che}}, \ and\ \bibinfo {author} {\bibfnamefont {S.}~\bibnamefont
  {Prestel}},\ }\href {\doibase 10.1103/PhysRevD.105.114012} {\bibfield
  {journal} {\bibinfo  {journal} {Phys. Rev. D}\ }\textbf {\bibinfo {volume}
  {105}},\ \bibinfo {pages} {114012} (\bibinfo {year} {2022})},\ \Eprint
  {http://arxiv.org/abs/2110.05964} {arXiv:2110.05964 [hep-ph]} \BibitemShut
  {NoStop}%
\bibitem [{\citenamefont {Ferrario~Ravasio}\ \emph {et~al.}(2023)\citenamefont
  {Ferrario~Ravasio}, \citenamefont {Hamilton}, \citenamefont {Karlberg},
  \citenamefont {Salam}, \citenamefont {Scyboz},\ and\ \citenamefont
  {Soyez}}]{FerrarioRavasio:2023kyg}%
  \BibitemOpen
  \bibfield  {author} {\bibinfo {author} {\bibfnamefont {S.}~\bibnamefont
  {Ferrario~Ravasio}}, \bibinfo {author} {\bibfnamefont {K.}~\bibnamefont
  {Hamilton}}, \bibinfo {author} {\bibfnamefont {A.}~\bibnamefont {Karlberg}},
  \bibinfo {author} {\bibfnamefont {G.~P.}\ \bibnamefont {Salam}}, \bibinfo
  {author} {\bibfnamefont {L.}~\bibnamefont {Scyboz}}, \ and\ \bibinfo {author}
  {\bibfnamefont {G.}~\bibnamefont {Soyez}},\ }\href {\doibase
  10.1103/PhysRevLett.131.161906} {\bibfield  {journal} {\bibinfo  {journal}
  {Phys. Rev. Lett.}\ }\textbf {\bibinfo {volume} {131}},\ \bibinfo {pages}
  {161906} (\bibinfo {year} {2023})},\ \Eprint
  {http://arxiv.org/abs/2307.11142} {arXiv:2307.11142 [hep-ph]} \BibitemShut
  {NoStop}%
\bibitem [{\citenamefont {van Beekveld}\ \emph {et~al.}(2025)\citenamefont {van
  Beekveld}, \citenamefont {Dasgupta}, \citenamefont {El-Menoufi},
  \citenamefont {Helliwell}, \citenamefont {Monni},\ and\ \citenamefont
  {Salam}}]{vanBeekveld:2024qxs}%
  \BibitemOpen
  \bibfield  {author} {\bibinfo {author} {\bibfnamefont {M.}~\bibnamefont {van
  Beekveld}}, \bibinfo {author} {\bibfnamefont {M.}~\bibnamefont {Dasgupta}},
  \bibinfo {author} {\bibfnamefont {B.~K.}\ \bibnamefont {El-Menoufi}},
  \bibinfo {author} {\bibfnamefont {J.}~\bibnamefont {Helliwell}}, \bibinfo
  {author} {\bibfnamefont {P.~F.}\ \bibnamefont {Monni}}, \ and\ \bibinfo
  {author} {\bibfnamefont {G.~P.}\ \bibnamefont {Salam}},\ }\href {\doibase
  10.1007/JHEP03(2025)209} {\bibfield  {journal} {\bibinfo  {journal} {JHEP}\
  }\textbf {\bibinfo {volume} {03}},\ \bibinfo {pages} {209} (\bibinfo {year}
  {2025})},\ \Eprint {http://arxiv.org/abs/2409.08316} {arXiv:2409.08316
  [hep-ph]} \BibitemShut {NoStop}%
\bibitem [{\citenamefont {Nagy}\ and\ \citenamefont
  {Soper}(2008)}]{Nagy:2008eq}%
  \BibitemOpen
  \bibfield  {author} {\bibinfo {author} {\bibfnamefont {Z.}~\bibnamefont
  {Nagy}}\ and\ \bibinfo {author} {\bibfnamefont {D.~E.}\ \bibnamefont
  {Soper}},\ }\href {\doibase 10.1088/1126-6708/2008/07/025} {\bibfield
  {journal} {\bibinfo  {journal} {JHEP}\ }\textbf {\bibinfo {volume} {07}},\
  \bibinfo {pages} {025} (\bibinfo {year} {2008})},\ \Eprint
  {http://arxiv.org/abs/0805.0216} {arXiv:0805.0216 [hep-ph]} \BibitemShut
  {NoStop}%
\bibitem [{\citenamefont {Fischer}\ \emph {et~al.}(2017)\citenamefont
  {Fischer}, \citenamefont {Lifson},\ and\ \citenamefont
  {Skands}}]{Fischer:2017htu}%
  \BibitemOpen
  \bibfield  {author} {\bibinfo {author} {\bibfnamefont {N.}~\bibnamefont
  {Fischer}}, \bibinfo {author} {\bibfnamefont {A.}~\bibnamefont {Lifson}}, \
  and\ \bibinfo {author} {\bibfnamefont {P.}~\bibnamefont {Skands}},\ }\href
  {\doibase 10.1140/epjc/s10052-017-5306-7} {\bibfield  {journal} {\bibinfo
  {journal} {Eur. Phys. J. C}\ }\textbf {\bibinfo {volume} {77}},\ \bibinfo
  {pages} {719} (\bibinfo {year} {2017})},\ \Eprint
  {http://arxiv.org/abs/1708.01736} {arXiv:1708.01736 [hep-ph]} \BibitemShut
  {NoStop}%
\bibitem [{\citenamefont {Richardson}\ and\ \citenamefont
  {Webster}(2020)}]{Richardson:2018pvo}%
  \BibitemOpen
  \bibfield  {author} {\bibinfo {author} {\bibfnamefont {P.}~\bibnamefont
  {Richardson}}\ and\ \bibinfo {author} {\bibfnamefont {S.}~\bibnamefont
  {Webster}},\ }\href {\doibase 10.1140/epjc/s10052-019-7429-5} {\bibfield
  {journal} {\bibinfo  {journal} {Eur. Phys. J. C}\ }\textbf {\bibinfo {volume}
  {80}},\ \bibinfo {pages} {83} (\bibinfo {year} {2020})},\ \Eprint
  {http://arxiv.org/abs/1807.01955} {arXiv:1807.01955 [hep-ph]} \BibitemShut
  {NoStop}%
\bibitem [{\citenamefont {Karlberg}\ \emph {et~al.}(2021)\citenamefont
  {Karlberg}, \citenamefont {Salam}, \citenamefont {Scyboz},\ and\
  \citenamefont {Verheyen}}]{Karlberg:2021kwr}%
  \BibitemOpen
  \bibfield  {author} {\bibinfo {author} {\bibfnamefont {A.}~\bibnamefont
  {Karlberg}}, \bibinfo {author} {\bibfnamefont {G.~P.}\ \bibnamefont {Salam}},
  \bibinfo {author} {\bibfnamefont {L.}~\bibnamefont {Scyboz}}, \ and\ \bibinfo
  {author} {\bibfnamefont {R.}~\bibnamefont {Verheyen}},\ }\href {\doibase
  10.1140/epjc/s10052-021-09378-0} {\bibfield  {journal} {\bibinfo  {journal}
  {Eur. Phys. J. C}\ }\textbf {\bibinfo {volume} {81}},\ \bibinfo {pages} {681}
  (\bibinfo {year} {2021})},\ \Eprint {http://arxiv.org/abs/2103.16526}
  {arXiv:2103.16526 [hep-ph]} \BibitemShut {NoStop}%
\bibitem [{\citenamefont {Hamilton}\ \emph {et~al.}(2022)\citenamefont
  {Hamilton}, \citenamefont {Karlberg}, \citenamefont {Salam}, \citenamefont
  {Scyboz},\ and\ \citenamefont {Verheyen}}]{Hamilton:2021dyz}%
  \BibitemOpen
  \bibfield  {author} {\bibinfo {author} {\bibfnamefont {K.}~\bibnamefont
  {Hamilton}}, \bibinfo {author} {\bibfnamefont {A.}~\bibnamefont {Karlberg}},
  \bibinfo {author} {\bibfnamefont {G.~P.}\ \bibnamefont {Salam}}, \bibinfo
  {author} {\bibfnamefont {L.}~\bibnamefont {Scyboz}}, \ and\ \bibinfo {author}
  {\bibfnamefont {R.}~\bibnamefont {Verheyen}},\ }\href {\doibase
  10.1007/JHEP03(2022)193} {\bibfield  {journal} {\bibinfo  {journal} {JHEP}\
  }\textbf {\bibinfo {volume} {03}},\ \bibinfo {pages} {193} (\bibinfo {year}
  {2022})},\ \Eprint {http://arxiv.org/abs/2111.01161} {arXiv:2111.01161
  [hep-ph]} \BibitemShut {NoStop}%
\bibitem [{\citenamefont {Staelin}\ \emph {et~al.}(1993)\citenamefont
  {Staelin}, \citenamefont {Morgenthaler},\ and\ \citenamefont
  {Kong}}]{Staelin:1994}%
  \BibitemOpen
  \bibfield  {author} {\bibinfo {author} {\bibfnamefont {D.~H.}\ \bibnamefont
  {Staelin}}, \bibinfo {author} {\bibfnamefont {A.~W.}\ \bibnamefont
  {Morgenthaler}}, \ and\ \bibinfo {author} {\bibfnamefont {J.~A.}\
  \bibnamefont {Kong}},\ }\href@noop {} {\emph {\bibinfo {title}
  {Electromagnetic Waves}}}\ (\bibinfo  {publisher} {Pearson},\ \bibinfo {year}
  {1993})\BibitemShut {NoStop}%
\bibitem [{\citenamefont {Campbell}\ \emph {et~al.}(2025)\citenamefont
  {Campbell}, \citenamefont {H{\"o}che}, \citenamefont {Knobbe}, \citenamefont
  {Preuss},\ and\ \citenamefont {Reichelt}}]{Campbell:2025lrs}%
  \BibitemOpen
  \bibfield  {author} {\bibinfo {author} {\bibfnamefont {J.~M.}\ \bibnamefont
  {Campbell}}, \bibinfo {author} {\bibfnamefont {S.}~\bibnamefont {H{\"o}che}},
  \bibinfo {author} {\bibfnamefont {M.}~\bibnamefont {Knobbe}}, \bibinfo
  {author} {\bibfnamefont {C.~T.}\ \bibnamefont {Preuss}}, \ and\ \bibinfo
  {author} {\bibfnamefont {D.}~\bibnamefont {Reichelt}},\ }\href@noop {} {\
  (\bibinfo {year} {2025})},\ \Eprint {http://arxiv.org/abs/2505.10408}
  {arXiv:2505.10408 [hep-ph]} \BibitemShut {NoStop}%
\bibitem [{\citenamefont {Catani}\ and\ \citenamefont
  {Grazzini}(2000)}]{Catani:1999ss}%
  \BibitemOpen
  \bibfield  {author} {\bibinfo {author} {\bibfnamefont {S.}~\bibnamefont
  {Catani}}\ and\ \bibinfo {author} {\bibfnamefont {M.}~\bibnamefont
  {Grazzini}},\ }\href {\doibase 10.1016/S0550-3213(99)00778-6} {\bibfield
  {journal} {\bibinfo  {journal} {Nucl. Phys. B}\ }\textbf {\bibinfo {volume}
  {570}},\ \bibinfo {pages} {287} (\bibinfo {year} {2000})},\ \Eprint
  {http://arxiv.org/abs/hep-ph/9908523} {arXiv:hep-ph/9908523} \BibitemShut
  {NoStop}%
\bibitem [{\citenamefont {Sudakov}(1956)}]{Sudakov:1954sw}%
  \BibitemOpen
  \bibfield  {author} {\bibinfo {author} {\bibfnamefont {V.~V.}\ \bibnamefont
  {Sudakov}},\ }\href@noop {} {\bibfield  {journal} {\bibinfo  {journal} {Sov.
  Phys. JETP}\ }\textbf {\bibinfo {volume} {3}},\ \bibinfo {pages} {65}
  (\bibinfo {year} {1956})}\BibitemShut {NoStop}%
\bibitem [{\citenamefont {Kleiss}\ and\ \citenamefont
  {Stirling}(1985)}]{Kleiss:1985yh}%
  \BibitemOpen
  \bibfield  {author} {\bibinfo {author} {\bibfnamefont {R.}~\bibnamefont
  {Kleiss}}\ and\ \bibinfo {author} {\bibfnamefont {W.~J.}\ \bibnamefont
  {Stirling}},\ }\href {\doibase 10.1016/0550-3213(85)90285-8} {\bibfield
  {journal} {\bibinfo  {journal} {Nucl. Phys. B}\ }\textbf {\bibinfo {volume}
  {262}},\ \bibinfo {pages} {235} (\bibinfo {year} {1985})}\BibitemShut
  {NoStop}%
\bibitem [{\citenamefont {Somogyi}\ \emph {et~al.}(2005)\citenamefont
  {Somogyi}, \citenamefont {Trocsanyi},\ and\ \citenamefont
  {Del~Duca}}]{Somogyi:2005xz}%
  \BibitemOpen
  \bibfield  {author} {\bibinfo {author} {\bibfnamefont {G.}~\bibnamefont
  {Somogyi}}, \bibinfo {author} {\bibfnamefont {Z.}~\bibnamefont {Trocsanyi}},
  \ and\ \bibinfo {author} {\bibfnamefont {V.}~\bibnamefont {Del~Duca}},\
  }\href {\doibase 10.1088/1126-6708/2005/06/024} {\bibfield  {journal}
  {\bibinfo  {journal} {JHEP}\ }\textbf {\bibinfo {volume} {06}},\ \bibinfo
  {pages} {024} (\bibinfo {year} {2005})},\ \Eprint
  {http://arxiv.org/abs/hep-ph/0502226} {arXiv:hep-ph/0502226} \BibitemShut
  {NoStop}%
\bibitem [{\citenamefont {Libby}\ and\ \citenamefont
  {Sterman}(1978)}]{Libby:1978bx}%
  \BibitemOpen
  \bibfield  {author} {\bibinfo {author} {\bibfnamefont {S.~B.}\ \bibnamefont
  {Libby}}\ and\ \bibinfo {author} {\bibfnamefont {G.~F.}\ \bibnamefont
  {Sterman}},\ }\href {\doibase 10.1103/PhysRevD.18.4737} {\bibfield  {journal}
  {\bibinfo  {journal} {Phys. Rev. D}\ }\textbf {\bibinfo {volume} {18}},\
  \bibinfo {pages} {4737} (\bibinfo {year} {1978})}\BibitemShut {NoStop}%
\bibitem [{\citenamefont {Ellis}\ \emph {et~al.}(1978)\citenamefont {Ellis},
  \citenamefont {Georgi}, \citenamefont {Machacek}, \citenamefont {Politzer},\
  and\ \citenamefont {Ross}}]{Ellis:1978sf}%
  \BibitemOpen
  \bibfield  {author} {\bibinfo {author} {\bibfnamefont {R.~K.}\ \bibnamefont
  {Ellis}}, \bibinfo {author} {\bibfnamefont {H.}~\bibnamefont {Georgi}},
  \bibinfo {author} {\bibfnamefont {M.}~\bibnamefont {Machacek}}, \bibinfo
  {author} {\bibfnamefont {H.~D.}\ \bibnamefont {Politzer}}, \ and\ \bibinfo
  {author} {\bibfnamefont {G.~G.}\ \bibnamefont {Ross}},\ }\href {\doibase
  10.1016/0370-2693(78)90023-0} {\bibfield  {journal} {\bibinfo  {journal}
  {Phys. Lett. B}\ }\textbf {\bibinfo {volume} {78}},\ \bibinfo {pages} {281}
  (\bibinfo {year} {1978})}\BibitemShut {NoStop}%
\bibitem [{\citenamefont {Ellis}\ \emph {et~al.}(1979)\citenamefont {Ellis},
  \citenamefont {Georgi}, \citenamefont {Machacek}, \citenamefont {Politzer},\
  and\ \citenamefont {Ross}}]{Ellis:1978ty}%
  \BibitemOpen
  \bibfield  {author} {\bibinfo {author} {\bibfnamefont {R.~K.}\ \bibnamefont
  {Ellis}}, \bibinfo {author} {\bibfnamefont {H.}~\bibnamefont {Georgi}},
  \bibinfo {author} {\bibfnamefont {M.}~\bibnamefont {Machacek}}, \bibinfo
  {author} {\bibfnamefont {H.~D.}\ \bibnamefont {Politzer}}, \ and\ \bibinfo
  {author} {\bibfnamefont {G.~G.}\ \bibnamefont {Ross}},\ }\href {\doibase
  10.1016/0550-3213(79)90105-6} {\bibfield  {journal} {\bibinfo  {journal}
  {Nucl. Phys. B}\ }\textbf {\bibinfo {volume} {152}},\ \bibinfo {pages} {285}
  (\bibinfo {year} {1979})}\BibitemShut {NoStop}%
\bibitem [{\citenamefont {Bassetto}\ \emph {et~al.}(1982)\citenamefont
  {Bassetto}, \citenamefont {Ciafaloni}, \citenamefont {Marchesini},\ and\
  \citenamefont {Mueller}}]{Bassetto:1982ma}%
  \BibitemOpen
  \bibfield  {author} {\bibinfo {author} {\bibfnamefont {A.}~\bibnamefont
  {Bassetto}}, \bibinfo {author} {\bibfnamefont {M.}~\bibnamefont {Ciafaloni}},
  \bibinfo {author} {\bibfnamefont {G.}~\bibnamefont {Marchesini}}, \ and\
  \bibinfo {author} {\bibfnamefont {A.~H.}\ \bibnamefont {Mueller}},\ }\href
  {\doibase 10.1016/0550-3213(82)90161-4} {\bibfield  {journal} {\bibinfo
  {journal} {Nucl. Phys. B}\ }\textbf {\bibinfo {volume} {207}},\ \bibinfo
  {pages} {189} (\bibinfo {year} {1982})}\BibitemShut {NoStop}%
\bibitem [{\citenamefont {Dokshitzer}\ \emph {et~al.}(1991)\citenamefont
  {Dokshitzer}, \citenamefont {Khoze}, \citenamefont {Mueller},\ and\
  \citenamefont {Troian}}]{Dokshitzer:1991wu}%
  \BibitemOpen
  \bibfield  {author} {\bibinfo {author} {\bibfnamefont {Y.~L.}\ \bibnamefont
  {Dokshitzer}}, \bibinfo {author} {\bibfnamefont {V.~A.}\ \bibnamefont
  {Khoze}}, \bibinfo {author} {\bibfnamefont {A.~H.}\ \bibnamefont {Mueller}},
  \ and\ \bibinfo {author} {\bibfnamefont {S.~I.}\ \bibnamefont {Troian}},\
  }\href@noop {} {\emph {\bibinfo {title} {{Basics of perturbative QCD}}}}\
  (\bibinfo  {publisher} {{Editions Fronti\`eres, Gif-sur-Yvettes}},\ \bibinfo
  {year} {1991})\BibitemShut {NoStop}%
\bibitem [{\citenamefont {Frixione}\ and\ \citenamefont
  {Webber}(2002)}]{Frixione:2002ik}%
  \BibitemOpen
  \bibfield  {author} {\bibinfo {author} {\bibfnamefont {S.}~\bibnamefont
  {Frixione}}\ and\ \bibinfo {author} {\bibfnamefont {B.~R.}\ \bibnamefont
  {Webber}},\ }\href {\doibase 10.1088/1126-6708/2002/06/029} {\bibfield
  {journal} {\bibinfo  {journal} {JHEP}\ }\textbf {\bibinfo {volume} {06}},\
  \bibinfo {pages} {029} (\bibinfo {year} {2002})},\ \Eprint
  {http://arxiv.org/abs/hep-ph/0204244} {arXiv:hep-ph/0204244} \BibitemShut
  {NoStop}%
\bibitem [{\citenamefont {Nason}(2004)}]{Nason:2004rx}%
  \BibitemOpen
  \bibfield  {author} {\bibinfo {author} {\bibfnamefont {P.}~\bibnamefont
  {Nason}},\ }\href {\doibase 10.1088/1126-6708/2004/11/040} {\bibfield
  {journal} {\bibinfo  {journal} {JHEP}\ }\textbf {\bibinfo {volume} {11}},\
  \bibinfo {pages} {040} (\bibinfo {year} {2004})},\ \Eprint
  {http://arxiv.org/abs/hep-ph/0409146} {arXiv:hep-ph/0409146} \BibitemShut
  {NoStop}%
\bibitem [{\citenamefont {Frixione}\ \emph {et~al.}(2007)\citenamefont
  {Frixione}, \citenamefont {Nason},\ and\ \citenamefont
  {Oleari}}]{Frixione:2007vw}%
  \BibitemOpen
  \bibfield  {author} {\bibinfo {author} {\bibfnamefont {S.}~\bibnamefont
  {Frixione}}, \bibinfo {author} {\bibfnamefont {P.}~\bibnamefont {Nason}}, \
  and\ \bibinfo {author} {\bibfnamefont {C.}~\bibnamefont {Oleari}},\ }\href
  {\doibase 10.1088/1126-6708/2007/11/070} {\bibfield  {journal} {\bibinfo
  {journal} {JHEP}\ }\textbf {\bibinfo {volume} {11}},\ \bibinfo {pages} {070}
  (\bibinfo {year} {2007})},\ \Eprint {http://arxiv.org/abs/0709.2092}
  {arXiv:0709.2092 [hep-ph]} \BibitemShut {NoStop}%
\bibitem [{\citenamefont {Alioli}\ \emph {et~al.}(2010)\citenamefont {Alioli},
  \citenamefont {Nason}, \citenamefont {Oleari},\ and\ \citenamefont
  {Re}}]{Alioli:2010xd}%
  \BibitemOpen
  \bibfield  {author} {\bibinfo {author} {\bibfnamefont {S.}~\bibnamefont
  {Alioli}}, \bibinfo {author} {\bibfnamefont {P.}~\bibnamefont {Nason}},
  \bibinfo {author} {\bibfnamefont {C.}~\bibnamefont {Oleari}}, \ and\ \bibinfo
  {author} {\bibfnamefont {E.}~\bibnamefont {Re}},\ }\href {\doibase
  10.1007/JHEP06(2010)043} {\bibfield  {journal} {\bibinfo  {journal} {JHEP}\
  }\textbf {\bibinfo {volume} {06}},\ \bibinfo {pages} {043} (\bibinfo {year}
  {2010})},\ \Eprint {http://arxiv.org/abs/1002.2581} {arXiv:1002.2581
  [hep-ph]} \BibitemShut {NoStop}%
\bibitem [{\citenamefont {H{\"o}che}\ \emph {et~al.}(2011)\citenamefont
  {H{\"o}che}, \citenamefont {Krauss}, \citenamefont {Sch{\"o}nherr},\ and\
  \citenamefont {Siegert}}]{Hoche:2010pf}%
  \BibitemOpen
  \bibfield  {author} {\bibinfo {author} {\bibfnamefont {S.}~\bibnamefont
  {H{\"o}che}}, \bibinfo {author} {\bibfnamefont {F.}~\bibnamefont {Krauss}},
  \bibinfo {author} {\bibfnamefont {M.}~\bibnamefont {Sch{\"o}nherr}}, \ and\
  \bibinfo {author} {\bibfnamefont {F.}~\bibnamefont {Siegert}},\ }\href
  {\doibase 10.1007/JHEP04(2011)024} {\bibfield  {journal} {\bibinfo  {journal}
  {JHEP}\ }\textbf {\bibinfo {volume} {04}},\ \bibinfo {pages} {024} (\bibinfo
  {year} {2011})},\ \Eprint {http://arxiv.org/abs/1008.5399} {arXiv:1008.5399
  [hep-ph]} \BibitemShut {NoStop}%
\bibitem [{\citenamefont {H{\"o}che}\ \emph {et~al.}(2012)\citenamefont
  {H{\"o}che}, \citenamefont {Krauss}, \citenamefont {Sch{\"o}nherr},\ and\
  \citenamefont {Siegert}}]{Hoeche:2011fd}%
  \BibitemOpen
  \bibfield  {author} {\bibinfo {author} {\bibfnamefont {S.}~\bibnamefont
  {H{\"o}che}}, \bibinfo {author} {\bibfnamefont {F.}~\bibnamefont {Krauss}},
  \bibinfo {author} {\bibfnamefont {M.}~\bibnamefont {Sch{\"o}nherr}}, \ and\
  \bibinfo {author} {\bibfnamefont {F.}~\bibnamefont {Siegert}},\ }\href
  {\doibase 10.1007/JHEP09(2012)049} {\bibfield  {journal} {\bibinfo  {journal}
  {JHEP}\ }\textbf {\bibinfo {volume} {09}},\ \bibinfo {pages} {049} (\bibinfo
  {year} {2012})},\ \Eprint {http://arxiv.org/abs/1111.1220} {arXiv:1111.1220
  [hep-ph]} \BibitemShut {NoStop}%
\bibitem [{\citenamefont {Alwall}\ \emph {et~al.}(2014)\citenamefont {Alwall},
  \citenamefont {Frederix}, \citenamefont {Frixione}, \citenamefont {Hirschi},
  \citenamefont {Maltoni}, \citenamefont {Mattelaer}, \citenamefont {Shao},
  \citenamefont {Stelzer}, \citenamefont {Torrielli},\ and\ \citenamefont
  {Zaro}}]{Alwall:2014hca}%
  \BibitemOpen
  \bibfield  {author} {\bibinfo {author} {\bibfnamefont {J.}~\bibnamefont
  {Alwall}}, \bibinfo {author} {\bibfnamefont {R.}~\bibnamefont {Frederix}},
  \bibinfo {author} {\bibfnamefont {S.}~\bibnamefont {Frixione}}, \bibinfo
  {author} {\bibfnamefont {V.}~\bibnamefont {Hirschi}}, \bibinfo {author}
  {\bibfnamefont {F.}~\bibnamefont {Maltoni}}, \bibinfo {author} {\bibfnamefont
  {O.}~\bibnamefont {Mattelaer}}, \bibinfo {author} {\bibfnamefont {H.~S.}\
  \bibnamefont {Shao}}, \bibinfo {author} {\bibfnamefont {T.}~\bibnamefont
  {Stelzer}}, \bibinfo {author} {\bibfnamefont {P.}~\bibnamefont {Torrielli}},
  \ and\ \bibinfo {author} {\bibfnamefont {M.}~\bibnamefont {Zaro}},\ }\href
  {\doibase 10.1007/JHEP07(2014)079} {\bibfield  {journal} {\bibinfo  {journal}
  {JHEP}\ }\textbf {\bibinfo {volume} {07}},\ \bibinfo {pages} {079} (\bibinfo
  {year} {2014})},\ \Eprint {http://arxiv.org/abs/1405.0301} {arXiv:1405.0301
  [hep-ph]} \BibitemShut {NoStop}%
\bibitem [{\citenamefont {Andr\'e}\ and\ \citenamefont
  {Sj{\"o}strand}(1998)}]{Andre:1997vh}%
  \BibitemOpen
  \bibfield  {author} {\bibinfo {author} {\bibfnamefont {J.}~\bibnamefont
  {Andr\'e}}\ and\ \bibinfo {author} {\bibfnamefont {T.}~\bibnamefont
  {Sj{\"o}strand}},\ }\href {\doibase 10.1103/PhysRevD.57.5767} {\bibfield
  {journal} {\bibinfo  {journal} {Phys. Rev. D}\ }\textbf {\bibinfo {volume}
  {57}},\ \bibinfo {pages} {5767} (\bibinfo {year} {1998})},\ \Eprint
  {http://arxiv.org/abs/hep-ph/9708390} {arXiv:hep-ph/9708390} \BibitemShut
  {NoStop}%
\bibitem [{\citenamefont {Catani}\ \emph {et~al.}(2001)\citenamefont {Catani},
  \citenamefont {Krauss}, \citenamefont {Kuhn},\ and\ \citenamefont
  {Webber}}]{Catani:2001cc}%
  \BibitemOpen
  \bibfield  {author} {\bibinfo {author} {\bibfnamefont {S.}~\bibnamefont
  {Catani}}, \bibinfo {author} {\bibfnamefont {F.}~\bibnamefont {Krauss}},
  \bibinfo {author} {\bibfnamefont {R.}~\bibnamefont {Kuhn}}, \ and\ \bibinfo
  {author} {\bibfnamefont {B.~R.}\ \bibnamefont {Webber}},\ }\href {\doibase
  10.1088/1126-6708/2001/11/063} {\bibfield  {journal} {\bibinfo  {journal}
  {JHEP}\ }\textbf {\bibinfo {volume} {11}},\ \bibinfo {pages} {063} (\bibinfo
  {year} {2001})},\ \Eprint {http://arxiv.org/abs/hep-ph/0109231}
  {arXiv:hep-ph/0109231} \BibitemShut {NoStop}%
\bibitem [{\citenamefont {L{\"o}nnblad}(2002)}]{Lonnblad:2001iq}%
  \BibitemOpen
  \bibfield  {author} {\bibinfo {author} {\bibfnamefont {L.}~\bibnamefont
  {L{\"o}nnblad}},\ }\href {\doibase 10.1088/1126-6708/2002/05/046} {\bibfield
  {journal} {\bibinfo  {journal} {JHEP}\ }\textbf {\bibinfo {volume} {05}},\
  \bibinfo {pages} {046} (\bibinfo {year} {2002})},\ \Eprint
  {http://arxiv.org/abs/hep-ph/0112284} {arXiv:hep-ph/0112284} \BibitemShut
  {NoStop}%
\bibitem [{\citenamefont {Mangano}\ \emph {et~al.}(2002)\citenamefont
  {Mangano}, \citenamefont {Moretti},\ and\ \citenamefont
  {Pittau}}]{Mangano:2001xp}%
  \BibitemOpen
  \bibfield  {author} {\bibinfo {author} {\bibfnamefont {M.~L.}\ \bibnamefont
  {Mangano}}, \bibinfo {author} {\bibfnamefont {M.}~\bibnamefont {Moretti}}, \
  and\ \bibinfo {author} {\bibfnamefont {R.}~\bibnamefont {Pittau}},\ }\href
  {\doibase 10.1016/S0550-3213(02)00249-3} {\bibfield  {journal} {\bibinfo
  {journal} {Nucl. Phys. B}\ }\textbf {\bibinfo {volume} {632}},\ \bibinfo
  {pages} {343} (\bibinfo {year} {2002})},\ \Eprint
  {http://arxiv.org/abs/hep-ph/0108069} {arXiv:hep-ph/0108069} \BibitemShut
  {NoStop}%
\bibitem [{\citenamefont {Krauss}(2002)}]{Krauss:2002up}%
  \BibitemOpen
  \bibfield  {author} {\bibinfo {author} {\bibfnamefont {F.}~\bibnamefont
  {Krauss}},\ }\href {\doibase 10.1088/1126-6708/2002/08/015} {\bibfield
  {journal} {\bibinfo  {journal} {JHEP}\ }\textbf {\bibinfo {volume} {08}},\
  \bibinfo {pages} {015} (\bibinfo {year} {2002})},\ \Eprint
  {http://arxiv.org/abs/hep-ph/0205283} {arXiv:hep-ph/0205283} \BibitemShut
  {NoStop}%
\bibitem [{\citenamefont {Lavesson}\ and\ \citenamefont
  {L{\"o}nnblad}(2008{\natexlab{a}})}]{Lavesson:2007uu}%
  \BibitemOpen
  \bibfield  {author} {\bibinfo {author} {\bibfnamefont {N.}~\bibnamefont
  {Lavesson}}\ and\ \bibinfo {author} {\bibfnamefont {L.}~\bibnamefont
  {L{\"o}nnblad}},\ }\href {\doibase 10.1088/1126-6708/2008/04/085} {\bibfield
  {journal} {\bibinfo  {journal} {JHEP}\ }\textbf {\bibinfo {volume} {04}},\
  \bibinfo {pages} {085} (\bibinfo {year} {2008}{\natexlab{a}})},\ \Eprint
  {http://arxiv.org/abs/0712.2966} {arXiv:0712.2966 [hep-ph]} \BibitemShut
  {NoStop}%
\bibitem [{\citenamefont {H{\"o}che}\ \emph {et~al.}(2009)\citenamefont
  {H{\"o}che}, \citenamefont {Krauss}, \citenamefont {Schumann},\ and\
  \citenamefont {Siegert}}]{Hoeche:2009rj}%
  \BibitemOpen
  \bibfield  {author} {\bibinfo {author} {\bibfnamefont {S.}~\bibnamefont
  {H{\"o}che}}, \bibinfo {author} {\bibfnamefont {F.}~\bibnamefont {Krauss}},
  \bibinfo {author} {\bibfnamefont {S.}~\bibnamefont {Schumann}}, \ and\
  \bibinfo {author} {\bibfnamefont {F.}~\bibnamefont {Siegert}},\ }\href
  {\doibase 10.1088/1126-6708/2009/05/053} {\bibfield  {journal} {\bibinfo
  {journal} {JHEP}\ }\textbf {\bibinfo {volume} {05}},\ \bibinfo {pages} {053}
  (\bibinfo {year} {2009})},\ \Eprint {http://arxiv.org/abs/0903.1219}
  {arXiv:0903.1219 [hep-ph]} \BibitemShut {NoStop}%
\bibitem [{\citenamefont {Hamilton}\ \emph {et~al.}(2009)\citenamefont
  {Hamilton}, \citenamefont {Richardson},\ and\ \citenamefont
  {Tully}}]{Hamilton:2009ne}%
  \BibitemOpen
  \bibfield  {author} {\bibinfo {author} {\bibfnamefont {K.}~\bibnamefont
  {Hamilton}}, \bibinfo {author} {\bibfnamefont {P.}~\bibnamefont
  {Richardson}}, \ and\ \bibinfo {author} {\bibfnamefont {J.}~\bibnamefont
  {Tully}},\ }\href {\doibase 10.1088/1126-6708/2009/11/038} {\bibfield
  {journal} {\bibinfo  {journal} {JHEP}\ }\textbf {\bibinfo {volume} {11}},\
  \bibinfo {pages} {038} (\bibinfo {year} {2009})},\ \Eprint
  {http://arxiv.org/abs/0905.3072} {arXiv:0905.3072 [hep-ph]} \BibitemShut
  {NoStop}%
\bibitem [{\citenamefont {L{\"o}nnblad}\ and\ \citenamefont
  {Prestel}(2012)}]{Lonnblad:2011xx}%
  \BibitemOpen
  \bibfield  {author} {\bibinfo {author} {\bibfnamefont {L.}~\bibnamefont
  {L{\"o}nnblad}}\ and\ \bibinfo {author} {\bibfnamefont {S.}~\bibnamefont
  {Prestel}},\ }\href {\doibase 10.1007/JHEP03(2012)019} {\bibfield  {journal}
  {\bibinfo  {journal} {JHEP}\ }\textbf {\bibinfo {volume} {03}},\ \bibinfo
  {pages} {019} (\bibinfo {year} {2012})},\ \Eprint
  {http://arxiv.org/abs/1109.4829} {arXiv:1109.4829 [hep-ph]} \BibitemShut
  {NoStop}%
\bibitem [{\citenamefont {L{\"o}nnblad}\ and\ \citenamefont
  {Prestel}(2013{\natexlab{a}})}]{Lonnblad:2012ng}%
  \BibitemOpen
  \bibfield  {author} {\bibinfo {author} {\bibfnamefont {L.}~\bibnamefont
  {L{\"o}nnblad}}\ and\ \bibinfo {author} {\bibfnamefont {S.}~\bibnamefont
  {Prestel}},\ }\href {\doibase 10.1007/JHEP02(2013)094} {\bibfield  {journal}
  {\bibinfo  {journal} {JHEP}\ }\textbf {\bibinfo {volume} {02}},\ \bibinfo
  {pages} {094} (\bibinfo {year} {2013}{\natexlab{a}})},\ \Eprint
  {http://arxiv.org/abs/1211.4827} {arXiv:1211.4827 [hep-ph]} \BibitemShut
  {NoStop}%
\bibitem [{\citenamefont {Pl{\"a}tzer}(2013)}]{Platzer:2012bs}%
  \BibitemOpen
  \bibfield  {author} {\bibinfo {author} {\bibfnamefont {S.}~\bibnamefont
  {Pl{\"a}tzer}},\ }\href {\doibase 10.1007/JHEP08(2013)114} {\bibfield
  {journal} {\bibinfo  {journal} {JHEP}\ }\textbf {\bibinfo {volume} {08}},\
  \bibinfo {pages} {114} (\bibinfo {year} {2013})},\ \Eprint
  {http://arxiv.org/abs/1211.5467} {arXiv:1211.5467 [hep-ph]} \BibitemShut
  {NoStop}%
\bibitem [{\citenamefont {H{\"o}che}\ \emph {et~al.}(2019)\citenamefont
  {H{\"o}che}, \citenamefont {Krause},\ and\ \citenamefont
  {Siegert}}]{Hoche:2019ncc}%
  \BibitemOpen
  \bibfield  {author} {\bibinfo {author} {\bibfnamefont {S.}~\bibnamefont
  {H{\"o}che}}, \bibinfo {author} {\bibfnamefont {J.}~\bibnamefont {Krause}}, \
  and\ \bibinfo {author} {\bibfnamefont {F.}~\bibnamefont {Siegert}},\ }\href
  {\doibase 10.1103/PhysRevD.100.014011} {\bibfield  {journal} {\bibinfo
  {journal} {Phys. Rev. D}\ }\textbf {\bibinfo {volume} {100}},\ \bibinfo
  {pages} {014011} (\bibinfo {year} {2019})},\ \Eprint
  {http://arxiv.org/abs/1904.09382} {arXiv:1904.09382 [hep-ph]} \BibitemShut
  {NoStop}%
\bibitem [{\citenamefont {Lavesson}\ and\ \citenamefont
  {L{\"o}nnblad}(2008{\natexlab{b}})}]{Lavesson:2008ah}%
  \BibitemOpen
  \bibfield  {author} {\bibinfo {author} {\bibfnamefont {N.}~\bibnamefont
  {Lavesson}}\ and\ \bibinfo {author} {\bibfnamefont {L.}~\bibnamefont
  {L{\"o}nnblad}},\ }\href {\doibase 10.1088/1126-6708/2008/12/070} {\bibfield
  {journal} {\bibinfo  {journal} {JHEP}\ }\textbf {\bibinfo {volume} {12}},\
  \bibinfo {pages} {070} (\bibinfo {year} {2008}{\natexlab{b}})},\ \Eprint
  {http://arxiv.org/abs/0811.2912} {arXiv:0811.2912 [hep-ph]} \BibitemShut
  {NoStop}%
\bibitem [{\citenamefont {Gehrmann}\ \emph {et~al.}(2013)\citenamefont
  {Gehrmann}, \citenamefont {H{\"o}che}, \citenamefont {Krauss}, \citenamefont
  {Sch{\"o}nherr},\ and\ \citenamefont {Siegert}}]{Gehrmann:2012yg}%
  \BibitemOpen
  \bibfield  {author} {\bibinfo {author} {\bibfnamefont {T.}~\bibnamefont
  {Gehrmann}}, \bibinfo {author} {\bibfnamefont {S.}~\bibnamefont {H{\"o}che}},
  \bibinfo {author} {\bibfnamefont {F.}~\bibnamefont {Krauss}}, \bibinfo
  {author} {\bibfnamefont {M.}~\bibnamefont {Sch{\"o}nherr}}, \ and\ \bibinfo
  {author} {\bibfnamefont {F.}~\bibnamefont {Siegert}},\ }\href {\doibase
  10.1007/JHEP01(2013)144} {\bibfield  {journal} {\bibinfo  {journal} {JHEP}\
  }\textbf {\bibinfo {volume} {01}},\ \bibinfo {pages} {144} (\bibinfo {year}
  {2013})},\ \Eprint {http://arxiv.org/abs/1207.5031} {arXiv:1207.5031
  [hep-ph]} \BibitemShut {NoStop}%
\bibitem [{\citenamefont {H{\"o}che}\ \emph {et~al.}(2013)\citenamefont
  {H{\"o}che}, \citenamefont {Krauss}, \citenamefont {Sch{\"o}nherr},\ and\
  \citenamefont {Siegert}}]{Hoeche:2012yf}%
  \BibitemOpen
  \bibfield  {author} {\bibinfo {author} {\bibfnamefont {S.}~\bibnamefont
  {H{\"o}che}}, \bibinfo {author} {\bibfnamefont {F.}~\bibnamefont {Krauss}},
  \bibinfo {author} {\bibfnamefont {M.}~\bibnamefont {Sch{\"o}nherr}}, \ and\
  \bibinfo {author} {\bibfnamefont {F.}~\bibnamefont {Siegert}},\ }\href
  {\doibase 10.1007/JHEP04(2013)027} {\bibfield  {journal} {\bibinfo  {journal}
  {JHEP}\ }\textbf {\bibinfo {volume} {04}},\ \bibinfo {pages} {027} (\bibinfo
  {year} {2013})},\ \Eprint {http://arxiv.org/abs/1207.5030} {arXiv:1207.5030
  [hep-ph]} \BibitemShut {NoStop}%
\bibitem [{\citenamefont {Frederix}\ and\ \citenamefont
  {Frixione}(2012)}]{Frederix:2012ps}%
  \BibitemOpen
  \bibfield  {author} {\bibinfo {author} {\bibfnamefont {R.}~\bibnamefont
  {Frederix}}\ and\ \bibinfo {author} {\bibfnamefont {S.}~\bibnamefont
  {Frixione}},\ }\href {\doibase 10.1007/JHEP12(2012)061} {\bibfield  {journal}
  {\bibinfo  {journal} {JHEP}\ }\textbf {\bibinfo {volume} {12}},\ \bibinfo
  {pages} {061} (\bibinfo {year} {2012})},\ \Eprint
  {http://arxiv.org/abs/1209.6215} {arXiv:1209.6215 [hep-ph]} \BibitemShut
  {NoStop}%
\bibitem [{\citenamefont {L{\"o}nnblad}\ and\ \citenamefont
  {Prestel}(2013{\natexlab{b}})}]{Lonnblad:2012ix}%
  \BibitemOpen
  \bibfield  {author} {\bibinfo {author} {\bibfnamefont {L.}~\bibnamefont
  {L{\"o}nnblad}}\ and\ \bibinfo {author} {\bibfnamefont {S.}~\bibnamefont
  {Prestel}},\ }\href {\doibase 10.1007/JHEP03(2013)166} {\bibfield  {journal}
  {\bibinfo  {journal} {JHEP}\ }\textbf {\bibinfo {volume} {03}},\ \bibinfo
  {pages} {166} (\bibinfo {year} {2013}{\natexlab{b}})},\ \Eprint
  {http://arxiv.org/abs/1211.7278} {arXiv:1211.7278 [hep-ph]} \BibitemShut
  {NoStop}%
\bibitem [{\citenamefont {H{\"o}che}\ \emph
  {et~al.}(2025{\natexlab{b}})\citenamefont {H{\"o}che}, \citenamefont
  {Krauss}, \citenamefont {Meinzinger},\ and\ \citenamefont
  {Reichelt}}]{Hoche:2025gsb}%
  \BibitemOpen
  \bibfield  {author} {\bibinfo {author} {\bibfnamefont {S.}~\bibnamefont
  {H{\"o}che}}, \bibinfo {author} {\bibfnamefont {F.}~\bibnamefont {Krauss}},
  \bibinfo {author} {\bibfnamefont {P.}~\bibnamefont {Meinzinger}}, \ and\
  \bibinfo {author} {\bibfnamefont {D.}~\bibnamefont {Reichelt}},\ }\href@noop
  {} {\  (\bibinfo {year} {2025}{\natexlab{b}})},\ \Eprint
  {http://arxiv.org/abs/2507.22837} {arXiv:2507.22837 [hep-ph]} \BibitemShut
  {NoStop}%
\bibitem [{\citenamefont {Bassetto}\ \emph {et~al.}(1983)\citenamefont
  {Bassetto}, \citenamefont {Ciafaloni},\ and\ \citenamefont
  {Marchesini}}]{Bassetto:1983mvz}%
  \BibitemOpen
  \bibfield  {author} {\bibinfo {author} {\bibfnamefont {A.}~\bibnamefont
  {Bassetto}}, \bibinfo {author} {\bibfnamefont {M.}~\bibnamefont {Ciafaloni}},
  \ and\ \bibinfo {author} {\bibfnamefont {G.}~\bibnamefont {Marchesini}},\
  }\href {\doibase 10.1016/0370-1573(83)90083-2} {\bibfield  {journal}
  {\bibinfo  {journal} {Phys. Rept.}\ }\textbf {\bibinfo {volume} {100}},\
  \bibinfo {pages} {201} (\bibinfo {year} {1983})}\BibitemShut {NoStop}%
\bibitem [{\citenamefont {Catani}\ and\ \citenamefont
  {Seymour}(1997)}]{Catani:1996vz}%
  \BibitemOpen
  \bibfield  {author} {\bibinfo {author} {\bibfnamefont {S.}~\bibnamefont
  {Catani}}\ and\ \bibinfo {author} {\bibfnamefont {M.~H.}\ \bibnamefont
  {Seymour}},\ }\href {\doibase 10.1016/S0550-3213(96)00589-5} {\bibfield
  {journal} {\bibinfo  {journal} {Nucl. Phys. B}\ }\textbf {\bibinfo {volume}
  {485}},\ \bibinfo {pages} {291} (\bibinfo {year} {1997})},\ \bibinfo {note}
  {[Erratum: Nucl.Phys.B 510, 503--504 (1998)]},\ \Eprint
  {http://arxiv.org/abs/hep-ph/9605323} {arXiv:hep-ph/9605323} \BibitemShut
  {NoStop}%
\bibitem [{\citenamefont {'t~Hooft}(1974)}]{tHooft:1973alw}%
  \BibitemOpen
  \bibfield  {author} {\bibinfo {author} {\bibfnamefont {G.}~\bibnamefont
  {'t~Hooft}},\ }\href {\doibase 10.1016/0550-3213(74)90154-0} {\bibfield
  {journal} {\bibinfo  {journal} {Nucl. Phys. B}\ }\textbf {\bibinfo {volume}
  {72}},\ \bibinfo {pages} {461} (\bibinfo {year} {1974})}\BibitemShut
  {NoStop}%
\bibitem [{\citenamefont {Maltoni}\ \emph {et~al.}(2003)\citenamefont
  {Maltoni}, \citenamefont {Paul}, \citenamefont {Stelzer},\ and\ \citenamefont
  {Willenbrock}}]{Maltoni:2002mq}%
  \BibitemOpen
  \bibfield  {author} {\bibinfo {author} {\bibfnamefont {F.}~\bibnamefont
  {Maltoni}}, \bibinfo {author} {\bibfnamefont {K.}~\bibnamefont {Paul}},
  \bibinfo {author} {\bibfnamefont {T.}~\bibnamefont {Stelzer}}, \ and\
  \bibinfo {author} {\bibfnamefont {S.}~\bibnamefont {Willenbrock}},\ }\href
  {\doibase 10.1103/PhysRevD.67.014026} {\bibfield  {journal} {\bibinfo
  {journal} {Phys. Rev. D}\ }\textbf {\bibinfo {volume} {67}},\ \bibinfo
  {pages} {014026} (\bibinfo {year} {2003})},\ \Eprint
  {http://arxiv.org/abs/hep-ph/0209271} {arXiv:hep-ph/0209271} \BibitemShut
  {NoStop}%
\bibitem [{\citenamefont {Chen}\ \emph {et~al.}(2021)\citenamefont {Chen},
  \citenamefont {Moult},\ and\ \citenamefont {Zhu}}]{Chen:2020adz}%
  \BibitemOpen
  \bibfield  {author} {\bibinfo {author} {\bibfnamefont {H.}~\bibnamefont
  {Chen}}, \bibinfo {author} {\bibfnamefont {I.}~\bibnamefont {Moult}}, \ and\
  \bibinfo {author} {\bibfnamefont {H.~X.}\ \bibnamefont {Zhu}},\ }\href
  {\doibase 10.1103/PhysRevLett.126.112003} {\bibfield  {journal} {\bibinfo
  {journal} {Phys. Rev. Lett.}\ }\textbf {\bibinfo {volume} {126}},\ \bibinfo
  {pages} {112003} (\bibinfo {year} {2021})},\ \Eprint
  {http://arxiv.org/abs/2011.02492} {arXiv:2011.02492 [hep-ph]} \BibitemShut
  {NoStop}%
\bibitem [{\citenamefont {Ellis}\ \emph {et~al.}(1981)\citenamefont {Ellis},
  \citenamefont {Ross},\ and\ \citenamefont {Terrano}}]{Ellis:1980wv}%
  \BibitemOpen
  \bibfield  {author} {\bibinfo {author} {\bibfnamefont {R.~K.}\ \bibnamefont
  {Ellis}}, \bibinfo {author} {\bibfnamefont {D.~A.}\ \bibnamefont {Ross}}, \
  and\ \bibinfo {author} {\bibfnamefont {A.~E.}\ \bibnamefont {Terrano}},\
  }\href {\doibase 10.1016/0550-3213(81)90165-6} {\bibfield  {journal}
  {\bibinfo  {journal} {Nucl. Phys. B}\ }\textbf {\bibinfo {volume} {178}},\
  \bibinfo {pages} {421} (\bibinfo {year} {1981})}\BibitemShut {NoStop}%
\bibitem [{\citenamefont {Maxwell}(1865)}]{Maxwell:1865zz}%
  \BibitemOpen
  \bibfield  {author} {\bibinfo {author} {\bibfnamefont {J.~C.}\ \bibnamefont
  {Maxwell}},\ }\href {\doibase 10.1098/rstl.1865.0008} {\bibfield  {journal}
  {\bibinfo  {journal} {Phil. Trans. Roy. Soc. Lond.}\ }\textbf {\bibinfo
  {volume} {155}},\ \bibinfo {pages} {459} (\bibinfo {year}
  {1865})}\BibitemShut {NoStop}%
\bibitem [{\citenamefont {Komiske}\ \emph {et~al.}(2018)\citenamefont
  {Komiske}, \citenamefont {Metodiev},\ and\ \citenamefont
  {Thaler}}]{Komiske:2017aww}%
  \BibitemOpen
  \bibfield  {author} {\bibinfo {author} {\bibfnamefont {P.~T.}\ \bibnamefont
  {Komiske}}, \bibinfo {author} {\bibfnamefont {E.~M.}\ \bibnamefont
  {Metodiev}}, \ and\ \bibinfo {author} {\bibfnamefont {J.}~\bibnamefont
  {Thaler}},\ }\href {\doibase 10.1007/JHEP04(2018)013} {\bibfield  {journal}
  {\bibinfo  {journal} {JHEP}\ }\textbf {\bibinfo {volume} {04}},\ \bibinfo
  {pages} {013} (\bibinfo {year} {2018})},\ \Eprint
  {http://arxiv.org/abs/1712.07124} {arXiv:1712.07124 [hep-ph]} \BibitemShut
  {NoStop}%
\end{thebibliography}%
\end{document}